\documentclass[twocolumn,superscriptaddress,showpacs,prd,aps,amsmath,amssymb,nofootinbib]{revtex4-1}
\usepackage{natbib}
\usepackage{graphicx,color}
\usepackage{amsmath,amssymb}
\usepackage{verbatim}
\usepackage{float}
\usepackage{wasysym}
\usepackage{amssymb,graphicx}
\usepackage{epsfig}
\usepackage{psfrag}
\usepackage{dsfont}
\usepackage{amsfonts}
\usepackage{mathrsfs}
\usepackage{multirow}
\usepackage{times}
\usepackage{bm}
\usepackage{hyperref}
\usepackage{scalefnt}
\usepackage{xspace}
\hypersetup{
  colorlinks=true,        
  linkcolor=blue,         
  citecolor=cyan,         
}
\usepackage{blindtext, rotating}
\usepackage{pifont}
\usepackage[normalem]{ulem}   
\usepackage{grffile}
\graphicspath{{./}}

\newcommand{\beq}{\begin{equation}} 
\newcommand{\eeq}{\end{equation}} 
\newcommand{\beqn}{\begin{eqnarray}} 
\newcommand{\eeqn}{\end{eqnarray}}

\newcommand{\zD}{{\raise1.0ex\hbox{${}^{\ \circ}$}}\!\!\!\!\!D}
\newcommand{\alone}{{\raise0.5ex\hbox{${}^{\ 1}$}}\!\!\!\!\alpha}

\newcommand{\nalam}{\mathrel{\raise0.9ex\hbox{$^\lambda$}\mkern-14mu
\lower0.0ex\hbox{$\nabla$}}}

\newcommand{\zeroD}{{\raise1.0ex\hbox{${}^{\ \circ}$}}\!\!\!\!\!D}

\newcommand{\zLap}{{\raise1.0ex\hbox{${}^{\ \circ}$}}\!\!\!\!\Delta}
\newcommand{\zna}{{\raise1.0ex\hbox{${}^{\ \circ}$}}\!\!\!\!\!\nabla}
\newcommand{\zS}{{\raise1.0ex\hbox{${}^{\ \circ}$}}\!\!\!\!\!S}

\newcommand{\be}{\begin{equation}}
\newcommand{\ee}{\end{equation}}

\def\QEQ{{%
    \setbox0\hbox{$I$}%
    \rlap{\hbox to \wd0{\hss--\hss}}\box0
}}

\begin{document}
\title{General Relativistic Magnetohydrodynamic Simulations of
  Accretion Disks Around Tilted Binary Black Holes of Unequal Mass}
\author{Milton Ruiz}
\affiliation{Departamento de Astronom\'{\i}a y Astrof\'{\i}sica, Universitat de Val\'encia,
  Dr. Moliner 50, 46100, Burjassot (Val\`encia), Spain}
\affiliation{Department of Physics, University of Illinois at Urbana-Champaign, Urbana, IL 61801, USA}
\author{Antonios Tsokaros}
\affiliation{Department of Physics, University of Illinois at Urbana-Champaign, Urbana, IL 61801, USA}
\affiliation{National Center for Supercomputing Applications, University of Illinois at Urbana-Champaign, Urbana, IL 61801, USA}
\author{Stuart L. Shapiro}
\affiliation{Department of Physics, University of Illinois at Urbana-Champaign, Urbana, IL 61801, USA}
\affiliation{Department of Astronomy \& NCSA, University of Illinois at Urbana-Champaign, Urbana, IL 61801, USA}
\date{\today}

\begin{abstract}
  We perform general relativistic simulations of magnetized, accreting disks onto spinning black hole binaries with mass ratio
  $q\equiv M_{\rm 1,irr}/M_{\rm 2,irr} = 1$, $2$ and $4$. The magnitude of the individual dimensionless black hole spins are all
  $\chi= 0.26$ and lie either along the initial orbital plane or $45^\circ$ above it. We evolve these systems throughout the late
  inspiral, merger and postmerger phases to identify the impact of the black hole spins and the binary mass ratio on any jet and
  their electromagnetic (Poynting) signatures. We find that incipient jets are launched from both black holes regardless of the
  binary mass ratio and along the
  spin directions as long as the force-free parameter $B^2/(8\,\pi\rho_0)$ in the funnel and above their poles is larger than one.
  At large distances the two jets merge into a single one.  This effect likely prevents the electromagnetic detection of individual jets. 
  As the accretion rate reaches a quasistationary state during the late predecoupling phase, we observe a sudden amplification
  of the outgoing Poynting luminosity that depends on the binary mass ratio. Following merger, the sudden change in the direction
  of the spin of the black hole remnant with respect to the spins of its progenitors causes a reorientation of the jet. The
  remnant jet drives a single, high-velocity, outward narrow beam collimated by a tightly wound, helical magnetic field which, in
  turn, boosts the Poynting luminosity. This effect is nearly mass-ratio independent. During this process, a kink is produced
  in the magnetic field lines, confining the jet. The kink propagates along the jet but rapidly decays, leaving no memory of
  the spin-shift. Our results suggest that the merger of misaligned, low-spinning, black hole binary mergers in low-mass disks
  may not provide a viable scenario to explain
  X-shaped radio galaxies if other features are not taken into account. However, the  sudden  changes in the outgoing luminosity
  at merger may help to identify mergers of black holes in active galactic nuclei, shedding light on black hole growth mechanisms and the
  observed co-evolution of their host galaxies.
\end{abstract}
\maketitle

\section{Introduction}
\label{sec:intro}
Accreting black holes (BHs) are probably some of the most common astrophysical systems throughout the universe.
BH masses typically lie between a few solar masses to more than $10^9M_\odot$ for the most extreme supermassive objects
(SMBHs). On the low-mass end, the typical remnant of binary neutron star or black hole-neutron star mergers
is a highly-spinning BH surrounded by a torus of gas debris~(see~e.g.~\cite{Kyutoku:2021icp,Baiotti:2016qnr,Ruiz:2021gsv}).
On the high-mass end, there is strong observational evidence that most galaxies harbor SMBHs in their
centers~\cite{Richstone:1998ky,Gebhardt:2000fk} and hence it is expected that when two galaxies merge
a supermassive black hole binary (SMBHBH) with a separation $\lesssim 100\ \rm kpc$ is formed~(see~e.g.
\cite{DeRosa:2019myq}). Dynamical friction and star ejection bring the binary to pc-scale separation
and, simultaneously, the BHs carve out a low-density, inner cavity just outside their orbit, beyond which 
a circumbinary accretion disk forms~\cite{Milosavljevic:2004cg,1994ApJ...421..651A}.

As the accretion takes place, electromagnetic (EM) radiation is produced from tenuous hot plasma in the disk
and, in most cases, from the magnetically-dominated outgoing jet launched from the poles of the BHs and its
interactions with the  environment~(see e.g~\cite{Gold:2013zma,Ruiz:2016rai,Cattorini:2022tvx}). This radiation
may be detectable by multiple EM instruments, such as FERMI~\cite{Kocevski:2017liw}, the Event Horizon
Telescope~\cite{EventHorizonTelescope:2019ths}, PanSTARRS~\cite{2016arXiv161205560C}, the HST~\cite{hubblecite},
and JWST~\cite{2023arXiv230101779M}, to name just a few. Gravitational waves (GWs) from inspiralling BHBHs or even their BH-disk
remnants~\cite{Wessel:2020hvu,Tsokaros:2022hjk} have been, or are expected to be, detected either by ground- or
space-based GW observatories~(see e.g.~\cite{2021NatRP...3..344B,LIGOScientific:2021usb,Verbiest:2021kmt,2017arXiv170200786A}
and references therein). Therefore, the possibility of coincident detection of gravitational radiation with 
EM radiation from these systems make them prime sources for multimessenger astronomy. 
However, multimessenger observations of accreting BHs call for a detailed understanding of the environment surrounding
them, and simultaneously the identification of the ``smoking guns'' that can be used to distinguish these systems
from other EM sources. Theoretical work on accreting BHs to date has sought to identify characteristic EM
features that may accompany GW signals (see~e.g.~\cite{1973A&A....24..337S,Abramowicz1988,Ichimaru1977,                                                                                    Narayan:1996gp,Giacomazzo:2012iv,Noble:2012xz,Farris:2012,Farris:2013uqa,Gold:2013zma,                                                                                                     Gold:2014b,Kelly:2017,Bowen:2018,Khan:2018}).

BH-disk systems have been studied in great detail for several decades, progressively incorporating more detailed
physical phenomena. Many different disk models ranging from geometrically-thin, optically-thick to
geometrically-thick, optically-thin disks have been studied~(see~e.g. \cite{1973A&A....24..337S,Abramowicz1988,
  Ichimaru1977,Narayan:1996gp}). All these efforts have been highlighted
by the recent, direct imaging of the hot, luminous plasma near the event horizon of the  SMBHs at
the center of M87~\cite{EventHorizonTelescope:2019dse} and by Sgr A$^*$ in our own Galaxy~\cite{2022ApJ...930L..12E}. 
In the recently released, high-resolution picture of the Centaurus A (NGC 5128) central core, a SMBH and its accretion
disk remain invisible, while its kpc-long relativistic jet has an unexpected shape~\cite{EventHorizonTelescope:2021iqj}.
It has been suggested that this image may reveal the presence of a SMBHBH with a separation of~$\sim 1$ milliparsec
instead of a single SMBH. Notice that in all the above theoretical studies, including this one, the gravitational effects
from the disk are neglected because typically $M_{\rm disk}/M_{\rm BH}\lesssim 10^{-2}$. Here
$M_{\rm disk}$ and $M_{\rm BH}$ are the mass of the disk and BH, respectively. However, some isolated BHs or BHBHs
detectable by LISA may find themselves immersed in extended disks with masses comparable or greater than the BHs themselves.
This may be particularly true of stellar-mass BHs in active galactic nuclei (AGNs) and
quasars, or SMBHs in extended disks formed in nascent or merging galactic nuclei \cite{Tsokaros:2022hjk,Wessel:2020hvu}.
Recently, we have explored the evolution of self-gravitating disks around tilted, highly spinning BHs with
$M_{\rm disk}/M_{\rm BH}\sim 0.2$~\cite{Tsokaros:2022hjk}. We found that tidal torques from the
disk induce a BH spin precession which can induce a reorientation of the relativistic jet powered by such systems. Such
a jet may be observed by various EM instruments.

Studies of accretion onto BHBHs are still immature and there are still many open questions. In particular, there is no consensus
about the environment in the vicinity of the merging BHs. Theoretical efforts involving  BHBHs in circumbinary disks
that incorporate some degree of relativistic effects and magnetic fields have been launched~(see e.g.~\cite{Giacomazzo:2012iv,
  Noble:2012xz,Farris:2012,Farris:2013uqa,Gold:2013zma,Gold:2014b,Kelly:2017,Bowen:2018,Khan:2018} and references therein).
In~\cite{Gold:2013zma,Gold:2014b,Khan:2018}
we adopted ideal general relativistic magnetohydrodynamics (GRMHD)  to probe EM signatures from
magnetized, circumbinary accretion disks onto unequal mass, {\it nonspinning} BHBHs during the late predecoupling and
post-decoupling phases. In all cases we found that dual magnetically-driven jets are launched from the poles
of the BHs through the Blandford-Znajek mechanism~\cite{Blandford:1977}, whose outgoing Poynting luminosity
is~$\sim 0.1\%$ of the accretion power. This finding may explain X-ray emission in AGNs~\cite{2011A&A...530A..42C}, or
$\gamma$-ray bursts from stellar-mass BHs, if the disk lifetime is~$\sim\mathcal{O}(1)\ \rm s$. We also showed that these
systems might explain the event GW150914~\cite{LIGOScientific:2016vlm} and its EM counterpart GW150915-GBM reported by the
Fermi GBM~0.4 s following the GW observation~\cite{Connaughton:2016umz}.

The dynamical formation of minidisk structures around spinning BHBHs and their potential EM signatures in
full general relativity has been studied recently~\cite{Paschalidis:2021ntt,Bright:2022hnl} (see
also~\cite{Combi:2021xeh} for similar studies using an approximate metric for the spacetime evolution).
It has been suggested that, as the accretion rate exhibits a quasi-periodic behavior whose amplitude depends on
the minidisks, they may be used to estimate BH spins through EM observations in the near
future~\cite{Paschalidis:2021ntt,Bright:2022hnl}.

In this work, we extend our own previous studies to consider circumbinary disks around  BHBHs with misaligned spins
and mass ratios $q\equiv M_{\rm 1,irr} /M_{\rm 2,irr} = 1$, $2$ and $4$, starting near the end of the binary-disk
predecoupling epoch. Here $M_{\rm i,irr}$, with $i=1,2$, is the irreducible mass of the $i$th-BH.
The fluid is modeled using a $\Gamma$-law equation of state (EOS), $P=(\Gamma-1)\,\epsilon\,\rho_0$.
We set $\Gamma = 4/3$, appropriate for radiation pressure-dominated, optically thick-disks. The disk is
initially threaded by a pure poloidal magnetic field confined to its interior. The
magnitude of the individual dimensionless BH spin in all cases is $\chi\equiv S/M^2_{\rm i,irr} = 0.26$, consistent
with a stochastic accretion in AGNs~(see e.g.~\cite{King:2008au,2012ApJ...753...15N}), lying either
in the initial orbital plane or $45^\circ$ above it. Here $S$ is the magnitude of the spin angular momentum.
We evolve these systems throughout the late inspiral, merger and postmerger phases to identify their unique EM emission
features and to analyze the impact of the BHs' spin and the binary mass-ratio. In particular, we probe whether the
precession or spin-shift of the BH remnant with respect to the spins of its progenitors leaves an observable imprint
in the outgoing Poynting luminosity, in the profile of the surrounding medium, or  in the magnetically driven-jet.
Such an imprint may be used to characterize the spin of the merging BHBH or give new insights on the formation channels
of X-shaped radio galaxies~\cite{1984MNRAS.210..929L,Gopal-Krishna:2003bjt}
as well as on searches for such systems (see e.g.~\cite{2015ApJ...810L...6R}).

Once the accretion rate reaches a quasi-steady state during predecoupling, we find that a magnetically-driven
(dual) outflow inside a  helical magnetic funnel emerges from both BHs along the direction of each individual BH
spin, as long as the force-free parameter within the funnel $b^2/(2\rho_0)\equiv B^2/(8\,\pi\rho_0)\gtrsim 1$. Here $B$ is
the magnitude of the magnetic field as measured by a comoving observer and $\rho_0$ is the rest-mass density. However, at
large distance from the BHs, where $b^2/(2\,\rho_0)\lesssim 1$, both funnels merge, pointing in the direction of the total
angular momentum of the system. These results are  consistent with those recently reported in~\cite{Cattorini:2022tvx},
where misaligned spinning BHBHs, with spin parameter $\chi=0.6$ and spin lying $45^\circ$ below the initial orbital plane
were considered immersed in a cloud of magnetized matter. These studies indicate that EM signatures from individual
jets may not be detectable as suggested previously (see~e.g.~\cite{Palenzuela:2010xn}).

During the predecoupling phase, the accretion rate exhibits a quasi-periodic behavior with a dominant
frequency $2\,\pi\,f_{\rm c}\simeq (3/4)\Omega_{\rm BHBH}$ regardless of the mass ratio. Here
$\Omega_{\rm BHBH}$ is the average orbital frequency of the system. Following the first $\sim 0.14 -
0.36 (M/10^6M_\odot)(1+z)\,\rm h$ [or $\sim 5 -14 (M/10M_\odot)(1+z)\,\rm ms$], we observe that, depending on the mass
ratio, the outgoing Poynting luminosity is significantly boosted. Consistent with~\cite{Gold:2014b}, we also observe
a new, though rather moderate boost in the luminosity during the re-brightening/afterglow phase that is nearly
mass-ratio independent. These sudden changes from the jet and from the disk can alter the EM emission and may be
used to distinguish BHBH in AGNs from single accreting BHs based on jet morphology 
\cite{Gold:2014b,Shapiro:2013qsa}.

During and after merger we track the magnetic field lines emanating from the BH apparent horizons and observe a slight
kink perturbation propagating along the funnel. This kink can be attributed to the spin-shift of the BH remnant with
respect to the spin of its progenitors. However, this perturbation is quickly damped in $\Delta t\sim$ few
$M$, leaving no memory of the reorientation of the spin. It has been suggested that the spin of SMBHs
is low if accretion is stochastic, which naturally reduces their spin (see e.g.
\cite{King:2005mv,King:2008au,2009ApJ...697L.141W,2012ApJ...753...15N,Lodato:2012yr}. Therefore, our results,
along with those in~\cite{Cattorini:2022tvx}, disfavor spin reorientations as a plausible scenario
to explain X-shaped radio galaxies, although the situation with massive, self-gravitating disks and/or higher spins may be different.
It is worth mentioning that the spins of SMBHs  inferred from the X-ray
observations of AGNs with relativistically broadened Fe K$\alpha$ lines imply that the spin magnitude of most
SMBHs should be $\chi \gtrsim 0.9$~(see~e.g.~\cite{Iwasawa:1996uh,Fabian:2002gj,Brenneman:2006hw,Reynolds:2013rva}) and
hence spin reorientations from highly spinning BHBHs may be a plausible scenario. In fact, for steady accretion from
a magnetized disk, the spin will relax to $\chi\gtrsim 0.9$~\cite{Gammie:2003qi}.
As pointed out in~\cite{Bogdanovic:2021aav}, GW observations from merging SMBHBHs are required to impose tight
constraints on their spin. 

The paper is structured as follows: In Sec.~\ref{sec:methods} we provide a qualitative overview of the evolution
of BHBHs in circumbinary disks to motivate  the choice of our binary configurations. We also summarize our numerical
methods employed  for generating initial data and provide a review of the methods used in our  numerical evolutions.
For further details, readers are referred to~\cite{Gold:2013zma,Gold:2014b,Khan:2018}. In Sec.~\ref{sec:results}
we describe our findings and discuss their astrophysical implications. We summarize our conclusions in Sec.
\ref{sec:discussion}. Geometrized units, where $G =1=c$, are adopted throughout unless stated
otherwise.
%
\begin{figure*}
\centering
\includegraphics[scale=0.27]{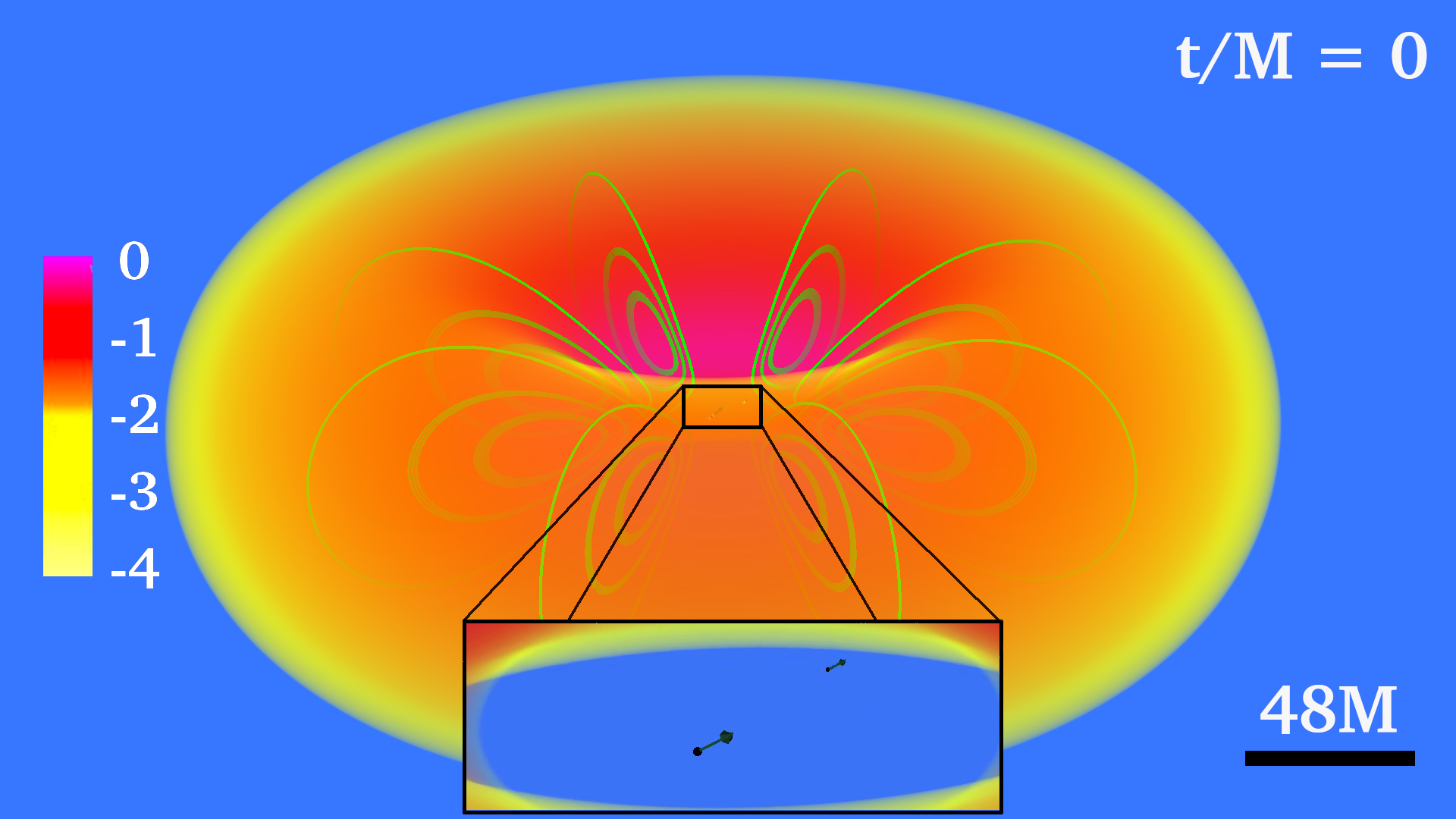}
\caption{Volume rendering of the rest-mass density normalized to its maximum value at $t=0$ for a BHBH with mass ratio
  $q=2$~(see Table~\ref{table:Iparamenters}). The disk is threaded by a pure poloidal magnetic field (green lines) confined
  to its interior. The inset highlights the coordinate position of the BH apparent horizons (black spheres)  and the
  direction of the individual BH spins (dark green arrow).\label{fig:q2_disk_ID}}
\end{figure*}
%
%
\section{Methods}
\label{sec:methods}
The dynamics of a BHBH  in a circumbinary disk can qualitatively be understood decomposing the evolution in three
phases (see~e.g.~\cite{Gold:2013zma,Khan:2018,Milosavljevic:2004cg,Shapiro:2009uy}):
i) the early inspiral (or predecoupling phase) during which the
evolution of the system is governed by a binary-disk angular momentum exchange mediated by tidal torques and an
effective viscosity. The matter follows the BHBH as it slowly inspirals;
ii) the very late inspiral, where the orbital separation shrinks due to GW radiation emission. The binary then
moves inward faster than the ambient matter (postdecoupling phase),
leaving the disk behind with a subsequent decrease in the accretion rate; and finally,
iii) the binary merger epoch, where subsequently the matter begins to flow into the ``partial'' hollow formed
during the inspiral (rebrightening/afterglow phase).

As discussed in~\cite{Gold:2013zma,Khan:2018}, the disk structure at decoupling has a key role in
determining the subsequent evolution and the EM emission. The inner part of the disk settles into
a quasi-equilibrium state on the viscous timescale,
\begin{eqnarray}
t_{\rm vis} \sim \frac{2}{3}\frac{R_{\rm in}^{2}}{\nu}\,,
\end{eqnarray}
where $\nu$ is the effective viscosity induced by MHD turbulence~\cite{Balbus:1998ja}, and
$R_{\rm in}$ is the radius of the disk inner edge. This viscosity  can be approximately
fit to an $\alpha$-disk law for an analytic estimate, i.e.
\begin{eqnarray}
  \nu(R) &=& \frac{2}{3}\,\alpha\,(P/\rho_0)\,\Omega_{\rm K}^{-1}\nonumber\\
  &\approx&\frac{2}{3}\,\alpha\,(R/M)^{1/2}(H/R)^2M\,,
\end{eqnarray}
where $H$ is the half-thickness of the disk, and $M$ the total mass of the BHBH. Notice that in the
above expression we have assumed vertical
hydrostatic equilibrium to derive an approximate relationship between $P/\rho_0$ and $H/R$
\cite{Shapiro:1983book}. In the early stage of the BHBH evolution, the GW inspiral timescale $t_{\rm GW}$
can be estimated as~\cite{Peters:1964}
\begin{equation}
t_{GW} \approx \frac{5}{16}\frac{a^{4}}{M^{3}\,\eta}\,,
\end{equation}
where $a$ is the binary separation and $\eta\equiv 4\,q/(1+q)^2$ is the reduced mass, with mass ratio
$q\equiv M_{\rm 1,irr}/M_{\rm 2,irr}>1$. Equating the viscous time scale and the GW inspiral time scale, and assuming that
$R_{\rm in}=\beta\, a$, yields the decoupling separation~\cite{Gold:2013zma,Khan:2018}
\begin{eqnarray}
  \frac{a_{d}}{M}\approx 11.5
  \left(\frac{\beta}{1.3}\right)^{3/5}\left(\frac{\eta}{1}\right)^{2/5}
  \left(\frac{\alpha}{0.1}\right)^{-2/5}\left(\frac{H/R}{0.3}\right)^{-4/5}
\label{eq:a_d}
\end{eqnarray}                
where the normalizations are based on the values found in our simulations
described below. This estimate suggests that the initial orbital separation
of BBH models in the predecoupling phase should be set to be larger than $\sim 11.5M$. In what follows,
we  consider spinning binaries on a quasi-circular orbit with mass ratio $q=1$,~$2$,~and~$4$.
Table~\ref{table:Iparamenters} summarizes the initial parameters of these systems.  Notice that
we choose an initial orbital separation  $\gtrsim 12.5M$. Consistent with the above estimate,
we observe that the binaries undergo $\sim 11$ quasi-circular obits before entering
the postdecoupling phase (see below). This allows the disk (at least the inner part) to relax,
giving rise to a quasi-stationary accretion flow. 
%
%
\subsection{Metric Initial data}
\begin{figure*}
\includegraphics[scale=0.50]{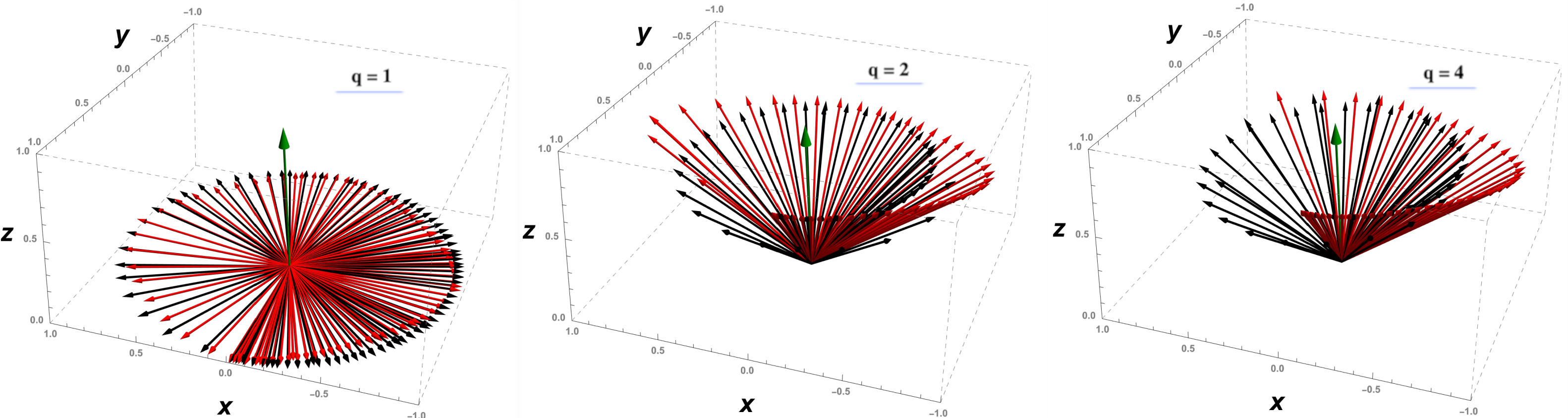}
\caption{
  Spin direction of each individual apparent horizon during inspiral (red arrows display the spin
  direction of the primary BH while those in black display the spin direction of the secondary) and the
  spin direction of the BH remnant following merger (green arrow) for all cases in Table~\ref{table:Iparamenters}.
  The arrows are plotted every $\Delta t\sim 40M$. Notice that the arrows indicate only the spin direction but
  not its magnitude. 
\label{fig:SpinBHs}}
\end{figure*}

Following~\cite{Khan:2018}, we use the {\tt TwoPunctures} code~\cite{Ansorg:2004ds} to build the spacetime
metric. We set the binaries on a quasi-circular orbit at a coordinate orbital separation $\gtrsim 12.5M$
(see~Table~\ref{table:Iparamenters}). Here $M$ is the ADM mass of the binary which is arbitrary.
To probe the effects of the mass ratio and the direction of the BHs' spin on the EM signatures and jet, we  consider
unequal mass BHBHs ($q=1$, $2$ and $4$) and set the magnitude of the individual dimensionless BH spin to
$\chi=0.26$, lying either in the initial orbital plane ($q=1$ case) or $45^\circ$ above it ($q=2$
and $q=4$ cases; see~inset in Fig.~\ref{fig:q2_disk_ID}), as in~\cite{Campanelli:2006fy}.
We consider BHs with a ``low-spin'' parameter to match those assuming that the accretion onto the BHs is stochastic
(see.~e.g.~\cite{King:2005mv,King:2008au,2009ApJ...697L.141W,2012ApJ...753...15N,Lodato:2012yr}).
It even has been argued that SMBH growth in an AGN occurs via sequences of randomly
oriented accretion disks and hence the BH spin parameter could be smaller ranging between $\chi\sim 0.1$
and~$\sim 0.3$~\cite{King:2008au}. As we describe below, our  configurations undergo $\gtrsim 14$ orbits before
merger, allowing the individual BH spins to preccess between half and a full cycle before merger (see
Fig.~\ref{fig:SpinBHs}). Details of the initial BHBH configurations are summarized in
Table~\ref{table:Iparamenters}.
%
%
\subsection{Fluid and Magnetic field Initial Data}
For the fluid we use the equilibrium solutions of a stationary disk around a single Kerr BH as
in e.g.~\cite{Farris:2011,DeVilliers:2003gr} with the same mass as the ADM mass of the BHBH
and adopting a $\Gamma$-law EOS, $P=(\Gamma-1)\,\epsilon\,\rho_0$.
In all models in Table~\ref{table:Iparamenters} we set $\Gamma=4/3$, which is appropriate for
radiation pressure-dominated, optically-thick disks~(see~e.g.~\cite{Farris:2009mt}).
The disk equator lies in the orbital x-y plane of the binary.
Following~\cite{Gold:2013zma,Khan:2018},
we choose the initial inner disk edge radius $R_{\rm in}=18M$ and the specific angular momentum
$l(R_{\rm in})=5.25M$. It is worth mentioning that our numerical studies of the BHBH-disk
in~\cite{Gold:2013zma}  found that by keeping the orbital separation fixed at $10M$ the accretion
rate for a disk with the above initial inner radius settles to a steady value by $t\sim 2000M$,
an indication that the inner disk edge has relaxed. In our cases, here, it is expected that the accretion
will settle earlier, since the BHs are closer to the disk, allowing the gravitational torques 
to plunder the inner layers of the disk easier than those above~\cite{Khan:2018}.

As in~\cite{Khan:2018,Gold:2013zma}, the accretion disk is initially threaded with
a dynamically unimportant, purely poloidal magnetic field that is confined to its
interior (see Fig.~\ref{fig:q2_disk_ID}). This field is generated by the vector potential
\begin{eqnarray}
A_{i} &=& \left(-\frac{y}{\varpi^{2}}{\delta^{x}}_{i} 
		+\frac{x}{\varpi^{2}}{\delta^{y}}_{i}\right)\,A_{\varphi}\,, \\
A_{\varphi} &=& A_B\,\varpi^{2}{\rm max}(P - P_{\rm cut},0)\,,
\end{eqnarray}
where $\varpi^{2} = x^2 + y^2$,  $A_B$, and $P_{\rm cut}$ are free
parameters. The cutoff pressure parameter $P_{\rm cut}$ confines the magnetic field
inside the disk to lie within the region where $P_{\rm gas}>P_{\rm cut}$.
We choose $P_{\rm cut}$ to be $1\%$ of the maximum pressure. The parameter $A_B$ determines the
strength of the magnetic field and can be characterized by the  magnetic-to-gas-pressure
ratio,  which we set to $P_{\rm mag}/P_{\rm   gas}\approx 0.01$.

\begin{figure}[ht]
\centering
\includegraphics[scale=0.55]{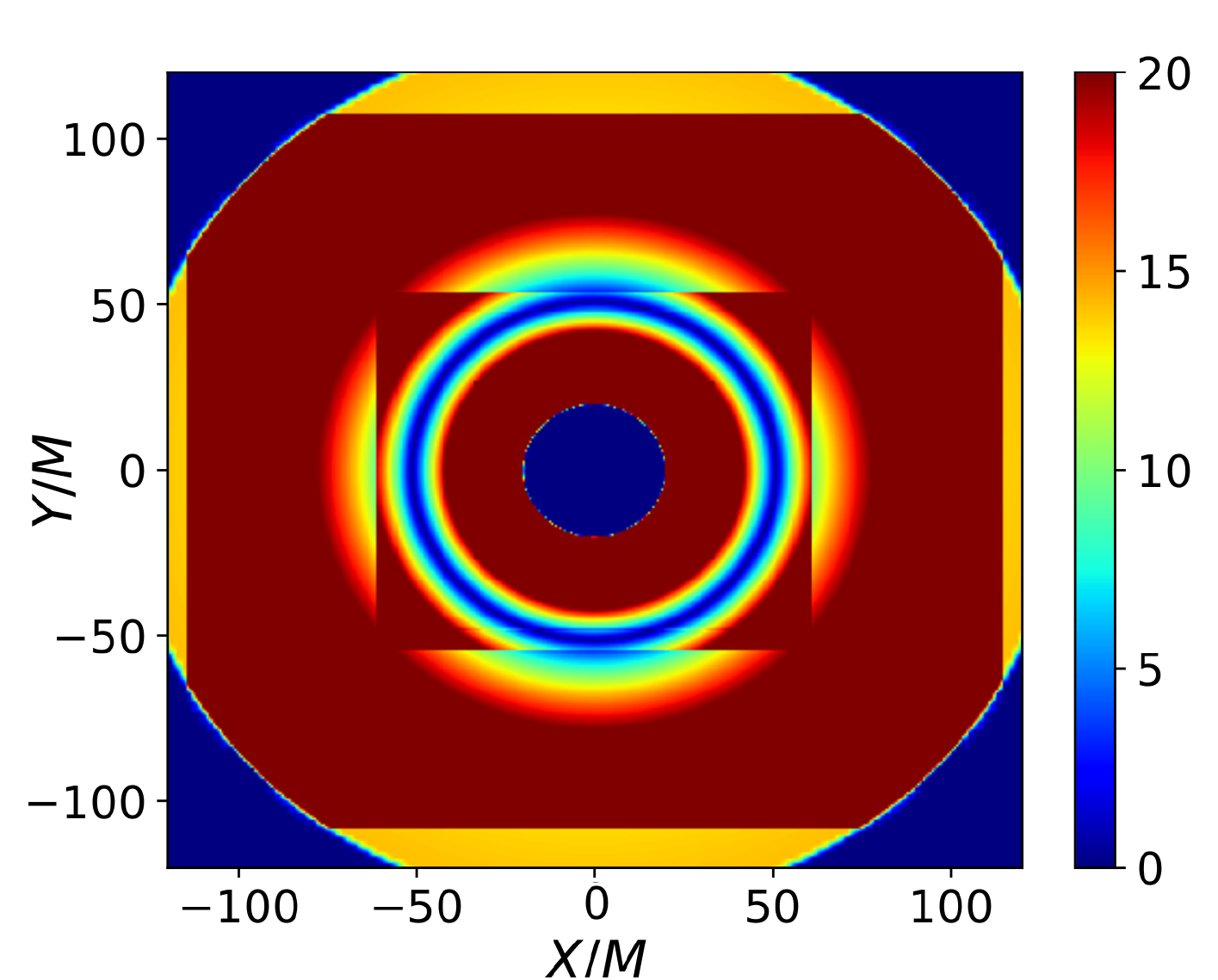}
\includegraphics[scale=0.55]{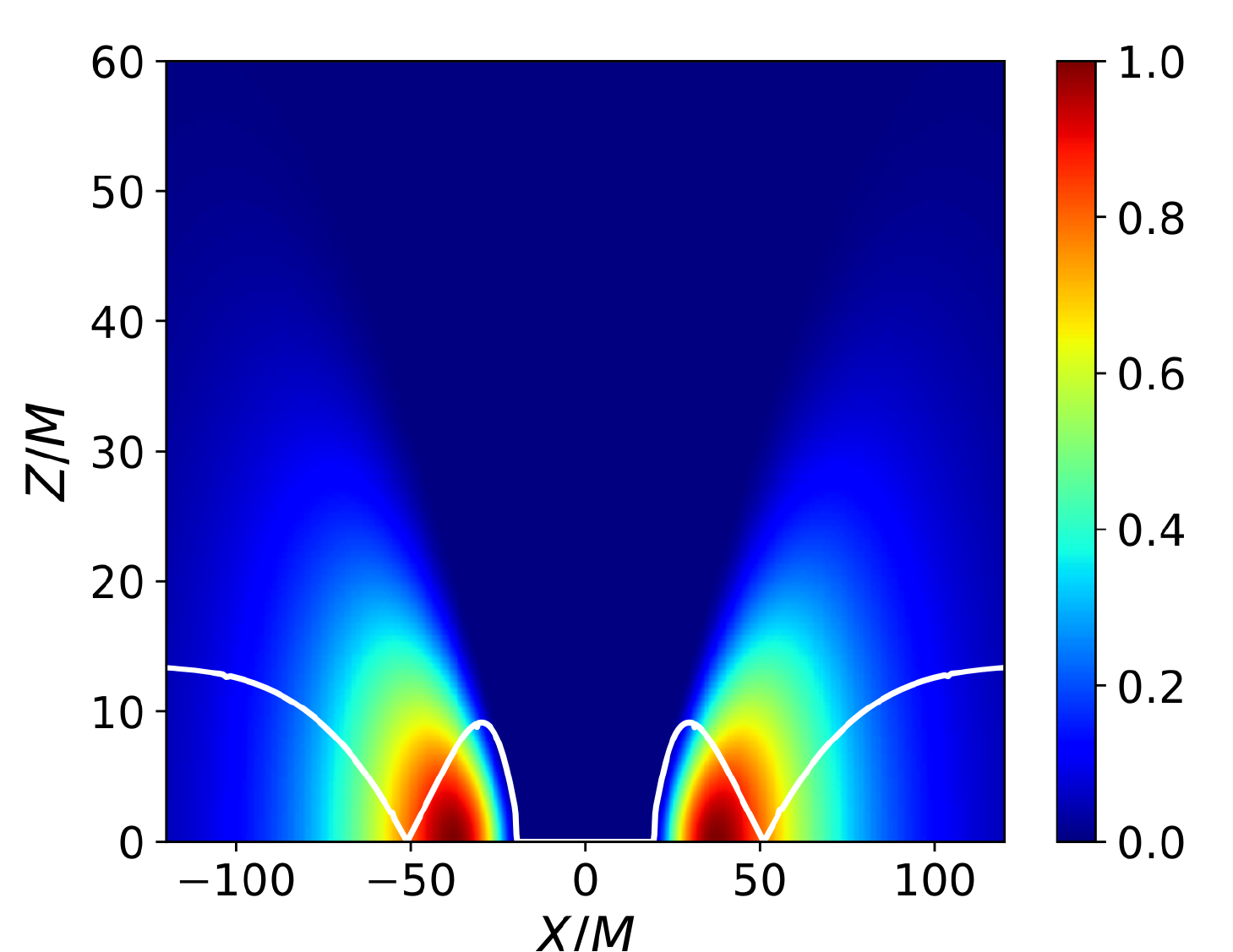}
\caption{Top panel: contours of the $\lambda_{\rm MRI}$-quality factor $Q = \lambda_{\rm MRI}/dx$ in the
  equatorial plane for case $q=1$. Notice that with the numerical resolution employed in our simulations the
  fastest growing MRI mode is resolved by $\gtrsim 20$ gridpoints over a large part of the disk except in a small region
  (blue ring) when the vertical component of the magnetic field changes sign. Bottom panel: initial rest-mass
  density contours (color coded) normalized to its maximum value on a  meridional slice along with the
  $\lambda_{\text{MRI}}/2$ (white solid line) demonstrating that the latter fits in the bulk of the disk.
  Similar results are found among all cases considered in Table~\ref{table:Iparamenters}.
\label{fig:mri}}
\end{figure}

Our disk models are unstable to the magnetorotational instability (MRI),
which induces an effective turbulent viscosity allowing angular momentum transport
and accretion to take place.  As pointed out in~\cite{Khan:2018,Gold:2013zma} in
our disk: i) we choose a rotation profile satisfying $\partial_{r}
\Omega < 0$, with $\Omega=u^{\phi}/u^{t}$ the fluid angular velocity~\cite{Duez:2006qe};
ii) in the bulk of the disk  we resolve the wavelength of the fastest growing MRI mode
$\lambda_{\text{MRI}}$ by $\gtrsim 20$ gridpoints~\cite{Gold:2013zma}, except for the
region near a radius where the magnetic field flips sign and becomes very small
(see top panel in Fig.~\ref{fig:mri}); and  iii) the strength of the magnetic field is
low enough for  $\lambda_{\text{MRI}}$  to fit inside the disk (see bottom panel in
Fig.~\ref{fig:mri}). 
%
\begin{center}
  \begin{table}[th]
    \caption{Two puncture initial data parameters for quasi-circular, spinning BHBH spacetimes. Columns
      display the case name, the mass ratio ($q\equiv M_{1,\rm{irr}}/M_{2,\rm{irr}}\geq 1$), defined as the
      ratio of the BH irreducible masses, the direction and magnitude
      of the dimensionless BH spin, the binary coordinate separation ($a/M$), the binary angular velocity
      $\Omega M$, and total ADM angular momentum $J/M^{2}$. Here $M$ is the ADM mass of the system.
      In all cases we set $S_x=0$ and hence the magnitude of the individual dimensionless BH spin is
      $\chi= S/M^2_{\rm irr} = 0.26$.
      \label{table:Iparamenters}}
    \begin{tabular}{ccccccc} 
      \hline
      \hline
       case& $q$   & $S_y/{M^2_{\rm i, irr}}$ & $S_z/{M^2_{\rm i, irr}}$  &$a/M$  & $\Omega M$ & $J/M^2$\\
      \hline
      q1    & 1     &  0.26                   & 0.0                        & 14.0  &  $1.74\times 10^{-2}$ & 1.07 \\
      q2    & 2     &  0.184                  & 0.184                      & 13.25 &  $1.89\times 10^{-2}$ & 1.03 \\
      q4    & 4     &  0.184                  & 0.184                      & 12.5  &  $2.04\times 10^{-2}$ & 0.78 \\
      \hline
    \end{tabular}
  \end{table}
\end{center}
%
%
%
%
\subsection{Numerical Setup}
\begin{paragraph}{\bf Evolution method:}
  We use the extensively tested {\tt Illinois GRMHD} code~\cite{Etienne:2010ui} which
  is embedded in the {\tt Cactus} infrastructure~\cite{CactusConfigs} and
  employs the {\tt Carpet} code \cite{Carpet,carpetweb} for moving-box mesh
  capability. For the metric evolution this code solves the
  Baumgarte--Shapiro--Shibata--Nakamura (BSSN) equations~\cite{Shibata:1995,Baumgarte:1998te},
  coupled to the puncture gauge conditions cast in first order form (see~Eq.~(14)-(16)
  in~\cite{Etienne:2007jg}). The code adopts fourth-order, centered, spatial differencing, except
  on shift advection terms, where fourth-order upwind differencing is used.
  For the damping coefficient $\eta_\beta$ appearing in the shift condition we use a
  spatially varying coefficient, as was done in the numerical simulations of the
  Numerical-Relativity-Analytical-Relativity (NRAR) collaboration~(see Table 3.~in~\cite{Hinder:2013oqa}).
  For the matter and magnetic field, the code solves the equations of
  ideal GRMHD in flux-conservative form (see Eqs.~21-24 in~\cite{Etienne:2007jg}) employing
  a high-resolution-shock-capturing scheme. To enforce the zero-divergence constraint of
  the magnetic field, we solve the magnetic induction equation using a vector
  potential formulation. To prevent spurious magnetic fields arising across
  refinement level we adopt the generalized Lorenz gauge~\cite{Etienne:2011re}.

  To verify the reliability of the BHBH evolution we implement a number of several local and global
  diagnostics as in~\cite{Gold:2013zma}. For example, we monitor the value of the $L_2$ norm
  of the normalized Hamiltonian and momentum constraint violations as introduced in Eqs. (40)-(41)
  in~\cite{Etienne:2007jg}, where the Laplacian operator is separated in its individual components
  when computing normalized norms. During the whole evolution the constraints remain $\lesssim 1\%$,
  peak at $\lesssim 2\%$ during merger, and finally settle back to $\lesssim 1\%$ after the BH-disk
  remnant reaches quasi-equilibrium.
\end{paragraph}
%
%
\begin{paragraph}
  {\bf Grid structure:}
  We use a set of three nested refinement boxes, with one centered on each BH and one set centeterd on the
  binary center of mass. In all cases, we use ten boxes centered on the primary BH while for the secondary
  one we use between $10$ to $12$ boxes depending on the mass ratio (i.e. depending on the initial size of
  the secondary BH's apparent horizon).
  The coarsest level has an outer boundary at $384M$ (see Table~\ref{table:Grids}).
  This hierarchical mesh structure allows us to resolve each individual apparent horizon by at
  least $\sim 26$ grid points across its radius and, simultaneously, to have enough resolution in the bulk of the
  disk to resolve~$\lambda_{\rm MRI}$ by $\gtrsim 20$ gridpoints (see~Fig.~\ref{fig:mri}).
\end{paragraph}
%
%
\section{Results}
\label{sec:results}
%
%
\begin{center}
  \begin{table*}[th]
    \caption{List of grid parameters for all models listed in Table~\ref{table:Iparamenters}. The computational
      mesh consists of three sets of nested moving grids, one centered on each BH and one on the binary
      center of mass, with the outer boundary at $380M$ in all cases. Here the columns indicate
      case label, the coarsest grid spacing $\Delta x_{\rm min}$, the number of refinement levels around each
      BHs and their half length. The grid spacing of all
      other levels is $\Delta x_{\rm max}/2^{n-1}$, with $n=1,\cdots, n_{\rm max}$, where $n_{\rm max}$ is
      the level number.
      \label{table:Grids}}
    \begin{tabular}{ccccccc} 
      \hline
      \hline
       case & $\Delta x_{\rm min}$   & Levels $\rm BH_1$ & levels $\rm BH_2$  &Grid hierarchy (in units of M) \\
      \hline
      q1    & $3.84M$      &  10           &  10            & $384M/2^{n-1}$, $n=1,\cdots,5$; $512M/2^{n}$, $n=6,\cdots,10$  \\
      q2    & $3.84M$      &  11           &  10            & $384M/2^{n-1}$, $n=1,\cdots,5$; $512M/2^{n}$, $n=6,\cdots,11$  \\
      q4    & $3.84M$      &  12           &  10            & $384M/2^{n-1}$, $n=1,\cdots,5$; $512M/2^{n}$, $n=6,\cdots,12$  \\
      \hline
    \end{tabular}
  \end{table*}
\end{center}
Our simulations are primarily designed to identify EM features associated with the jets that can be, along with gravitational radiation,
used to observationally infer some physical properties of BHBH, such as mass ratios, BH spin etc. In addition,
our studies are a step forward in addressing the question:\\
\indent {\it Is the reorientation of the spin direction of the BH remnant and associated jet, with respect to the
  spins of its progenitors, a viable mechanism to explain X-shaped radio galaxies?}\\

As discussed previously, we consider ``low-spin'' binaries consistent with a stochastic
accretion that naturally reduces the BH spin. We will explore high spinning BHBHs, consistent with
X-ray observations of AGNs with relativistically broadened Fe K$\alpha$ lines, in future investigations.
We also consider low-mass disks here with negligible self-gravity. We will treat high-mass disks with 
significant self-gravity in future studies as well.
%
%
%
\subsection{Accretion rates}
\label{sec:accretion}
In contrast to the nonspinning BHBH in our previous accretion disk studies~\cite{Gold:2013zma,Khan:2018}, where the
accretion takes place in the orbital plane, we observe that during the initial phase of the binary inspiral,
tidal torques strip matter from the inner disk's edge, which induces the formation of spiral arms that continuously
feed the BHs on a plane perpendicular to the direction of each individual spin. Similar behavior has been reported on
the misaligned spinning BHBHs immersed in  a magnetized cloud of matter recently reported in~\cite{Cattorini:2022tvx},
and in the misaligned, highly (extremal) spinning BHs surrounded by self-gravitating
disks~\cite{Tsokaros:2022hjk}.
%
%
\begin{figure*}
  \includegraphics[scale=0.25]{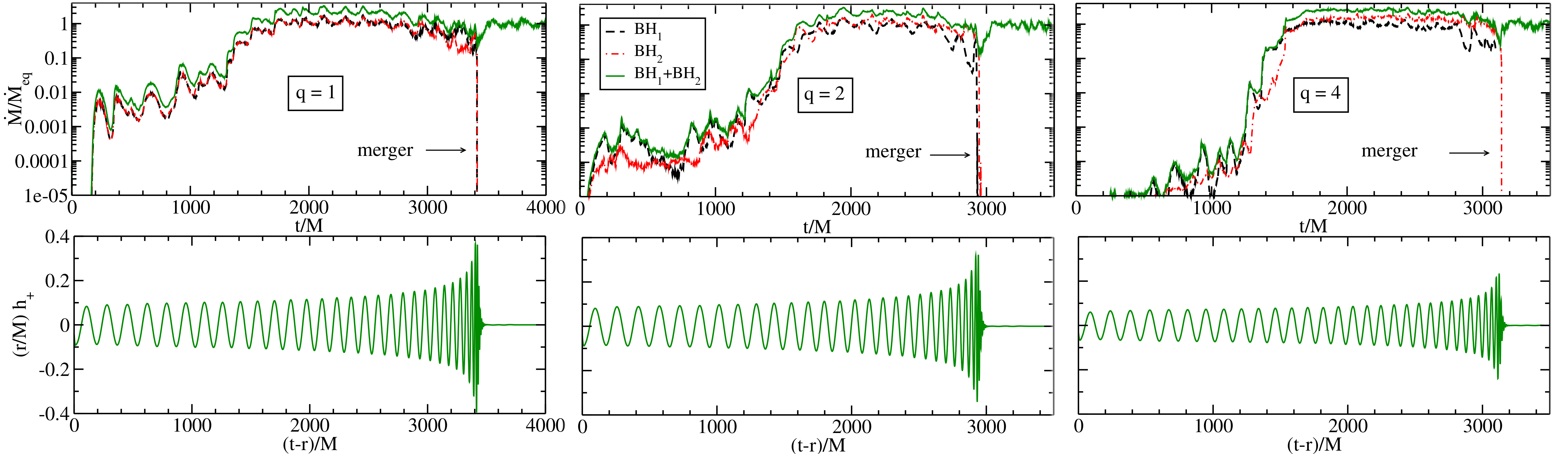}
  \caption{Top panel: Accretion rate $\dot{M}$ onto the primary BH (red, dashed line), onto its companion
    (black, solid line), and their sum (green, solid line) for all cases listed in
    Table~\ref{table:Iparamenters}. The normalization $\dot{M}_{\rm eq}$ is the time-averaged accretion rate onto
    the BH remnant over the last $\Delta t\sim 500M$ before the termination of our simulations.
    Bottom Panel:  GW strain $h^{22}_+$ (dominant mode) vs. retarded time.  
\label{fig:Mdot_GW}}
\end{figure*}

The top panel in Fig.~\ref{fig:Mdot_GW} displays the accretion rates $\dot{M}$ through the BH's
apparent horizons during the whole evolution, as introduced in  Eq.(A11) in~\cite{Farris:2009mt}.
During roughly the first $\sim 5$ quasi-circular orbits (or $\Delta t\sim 1500M$; see bottom panel
in~Fig.~\ref{fig:Mdot_GW}) we observe that when the mass ratio $q>1$, matter is more efficiently
swallowed by the primary BH than by the secondary, where a low-density, disk-like structure tends
to pile up around it periodically. This result is consistent with the studies in~\cite{Paschalidis:2021ntt,Gold:2014b}
showing that if the Hill sphere is well outside of the innermost stable circular orbit (ISCO) a
persistent disk-like structure (the so-called minidisks) can form around a BH. In our cases (see Table
\ref{table:Iparamenters}), the Hill sphere
can be estimated as $r_{\rm Hill}\equiv a\,(3\,q)^{-1/3}/2\sim 4M(a/11.5M)\,q^{-1/3}$. For the primary BH in
$q=4$ we find that $r_{\rm Hill}\sim 2.8M<r_{\rm ISCO}\sim 4M$. However for that secondary BH we find that
find that $r_{\rm Hill}\sim 2.8M>r_{\rm ISCO}\sim 1M$. For the primary BH in $q=2$ we find that
$r_{\rm Hill}\sim r_{\rm ISCO}\sim 3.5M$, while for the secondary one $r_{\rm Hill}\sim 2\,r_{\rm ISCO}\sim 3.5M$.
Finally, in $q=1$ we find that $r_{\rm Hill}\sim 2\,r_{\rm ISCO}\sim 5M$.  These Newtonian estimates explain
why we observe quasi-periodic modulations in the accretion rate (see top Panel in Fig.~\ref{fig:Mdot_GW})
in $q=1$ and in the secondary BH for the other two cases: there is enough room between $r_{\rm Hill}$
and for $r_{\rm ISCO}$ matter to wrap around the BHs, forming transient minidisks which in turn reduce 
the accretion. Note that persistent minidisks that form when $r_{\rm Hill}\gg r_{\rm ISCO}$ can 
suppress the accretion altogether~\cite{Paschalidis:2021ntt}. Thus, during the early inspiral the quasi-periodic behavior
in~$\dot{M}$ is dominated by the secondary BH.

Following the first $\sim 5$ quasi-circular orbits, we observe that, in all cases, $\dot{M}$ gradually reaches a
quasi-equilibrium state, and remains there for the next $\sim 5-6$ orbits ($\sim 1000M$), which indicates that the
system has settled into the quasi-stationary predecoupling phase. During this period, we compute the averaged accretion rate for
cases $q=2$ and $q=4$ and find that in the former the averaged accretion rate onto the primary BH is $\sim 16\%$
larger than the rate on its companion. However, for the latter case, the accretion onto the primary BH is $38\%$ larger.
This difference is expected, since the effective cross section of the secondary BH, which depends on its irreducible
mass, is smaller.
%
\begin{figure*}
\begin{center}
  \textbf{inspiral}\hspace{7cm}
  \textbf{postmerger}
  \par\medskip
 \begin{turn}{90}  
{\hspace{2.5cm}\bf q = 1}
\end{turn}
\includegraphics[scale=0.128]{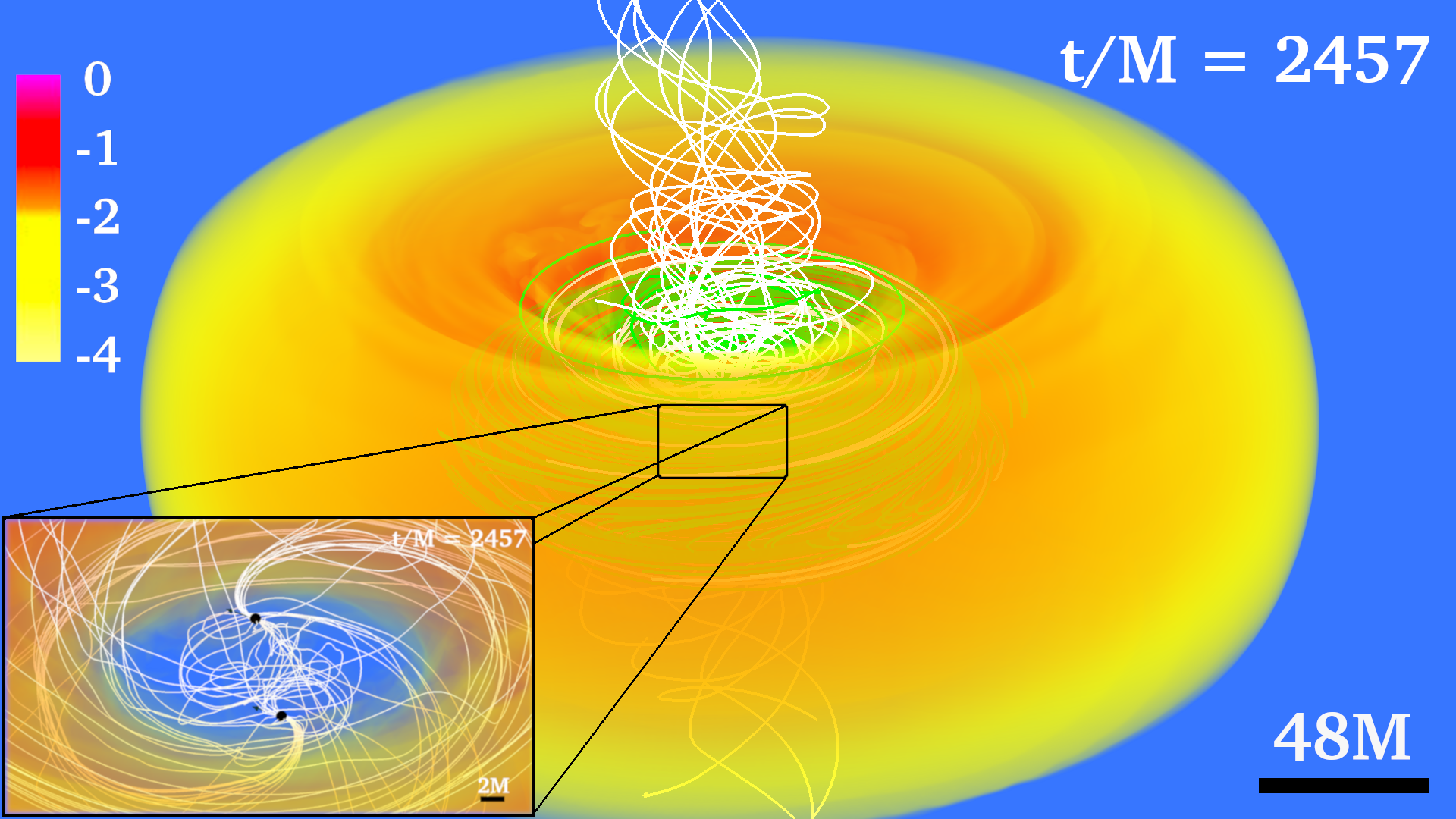}
\includegraphics[scale=0.128]{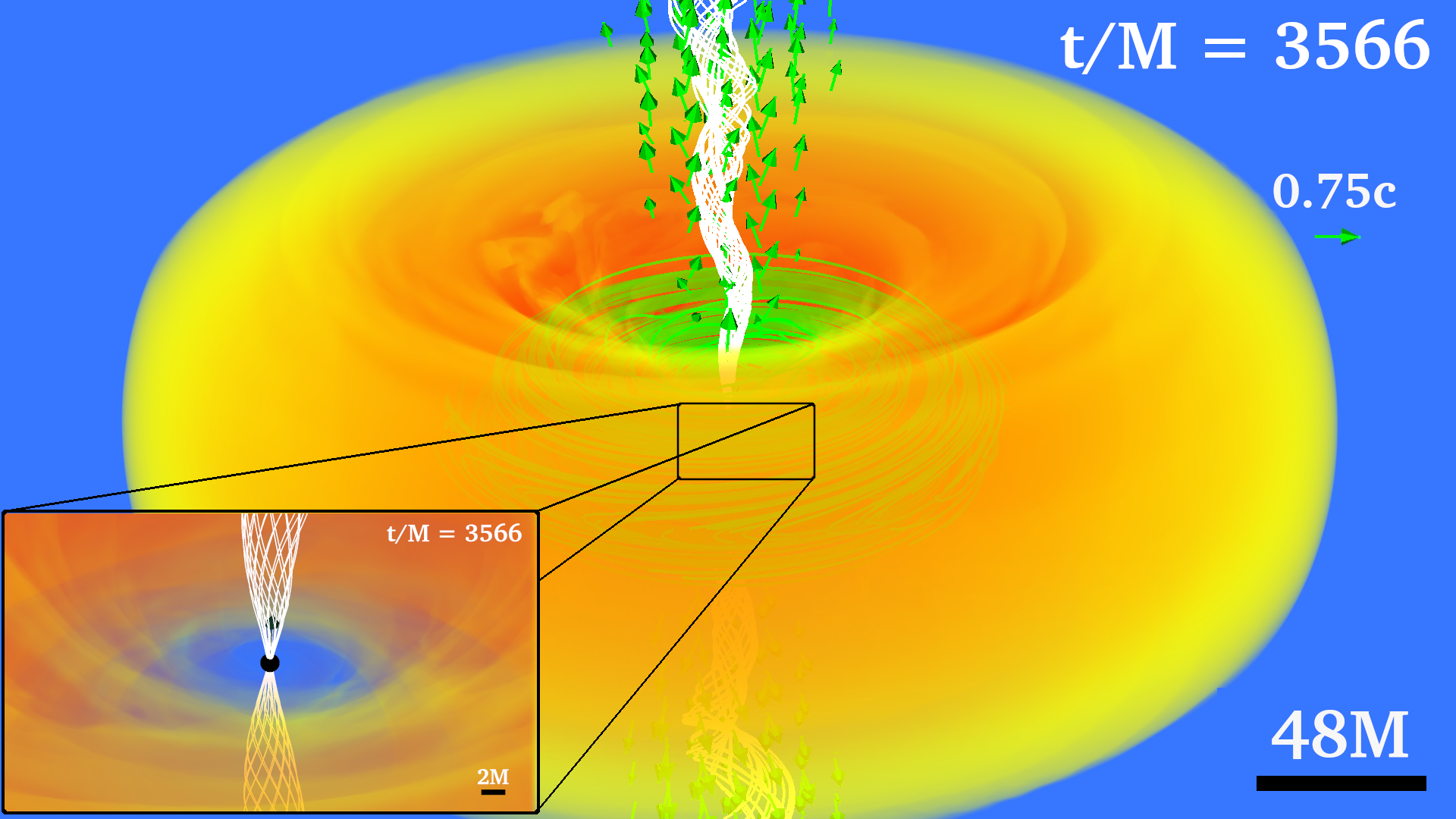}\\
\begin{turn}{90}  
{\hspace{2.5cm}\bf q = 2}
\end{turn}
\includegraphics[scale=0.128]{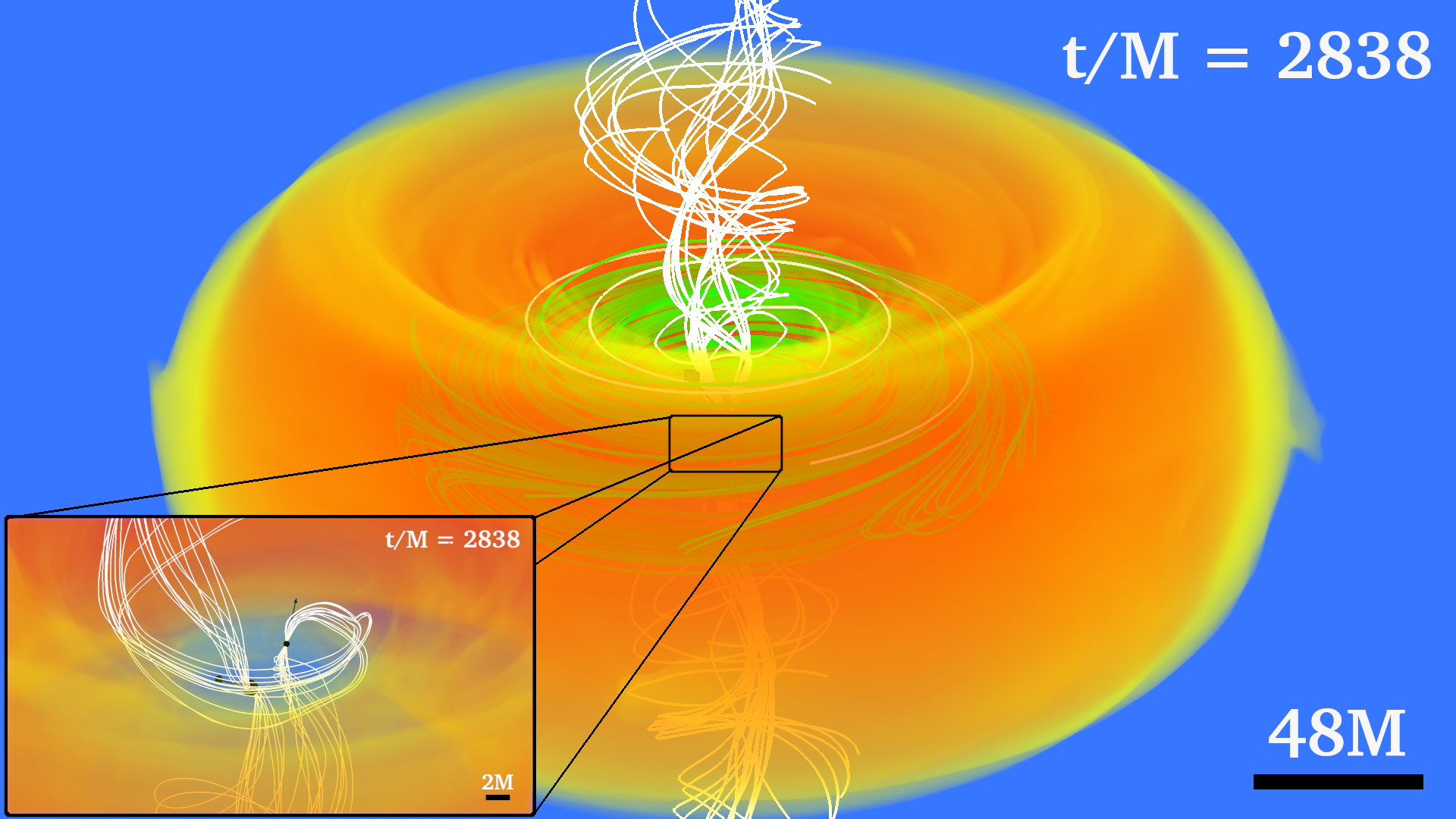}
\includegraphics[scale=0.128]{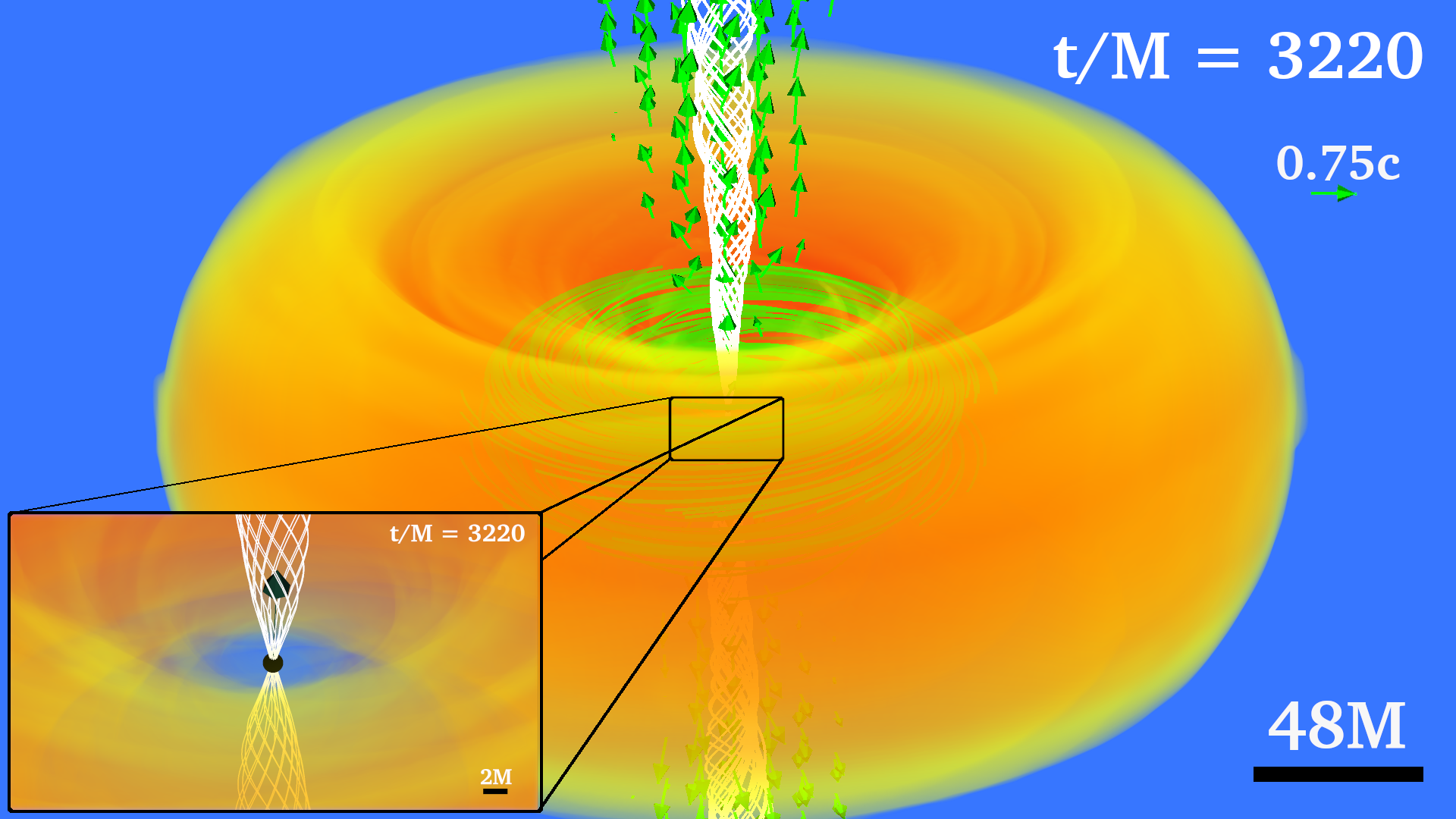}\\
\begin{turn}{90}
{\hspace{2.5cm}\bf q = 4}
\end{turn}
\includegraphics[scale=0.128]{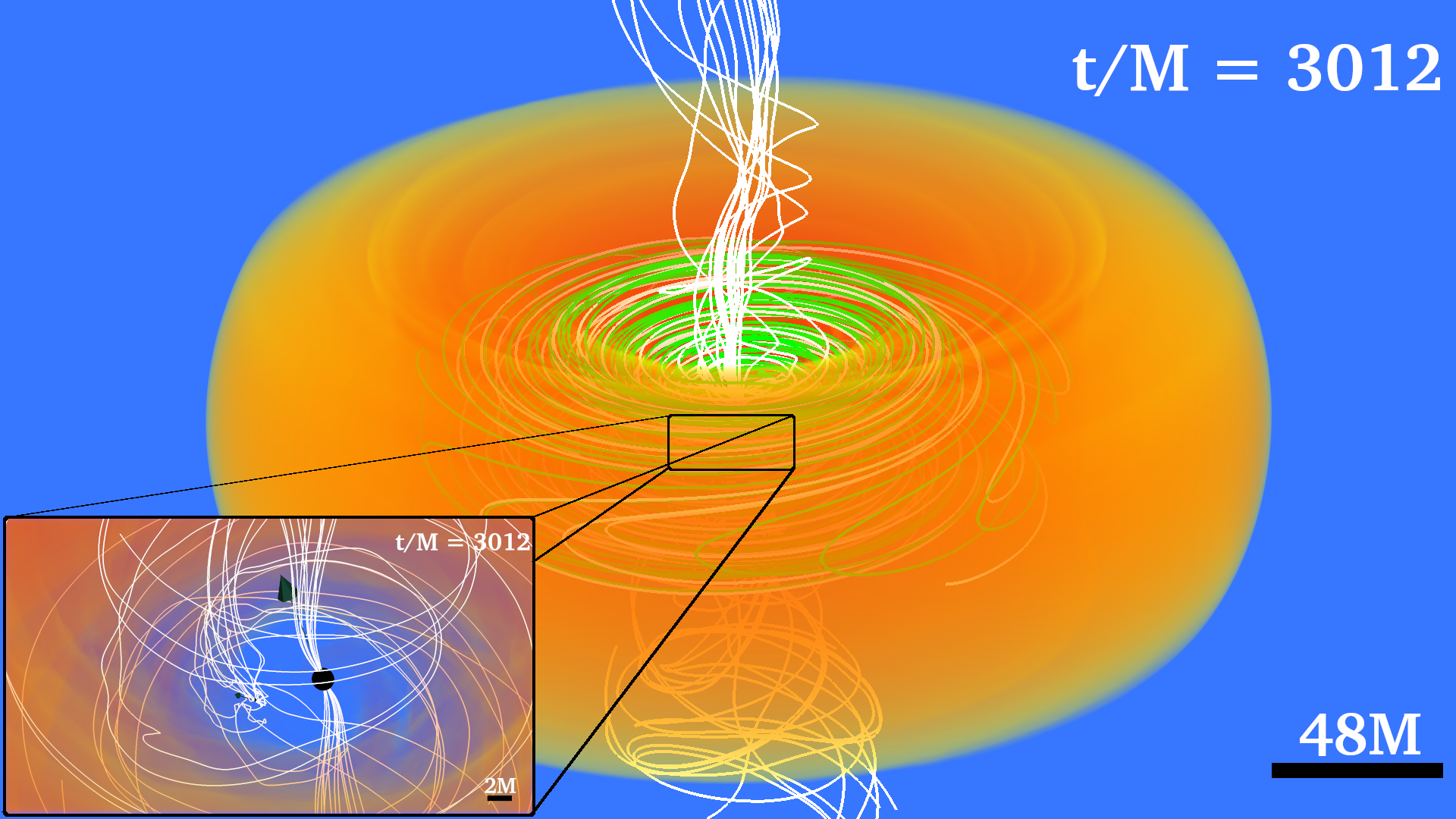}
\includegraphics[scale=0.128]{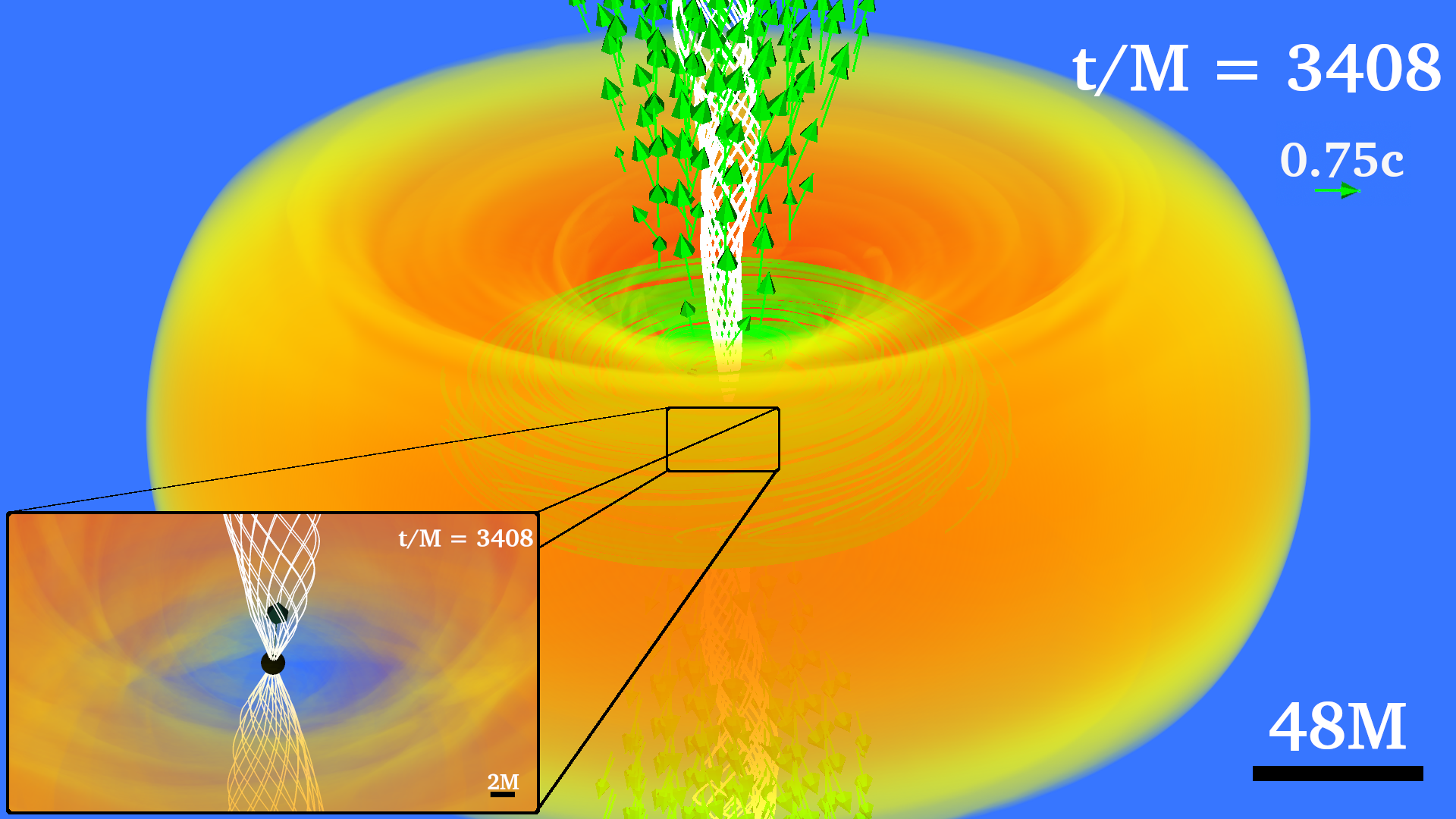}
\caption{Volume rendering of the rest-mass density profile,  normalized to the initial maximum density, during
  the predecoupling (left column) and postmerger rebrightening/afterglow phases (right column). Also shown are the magnetic
  field lines emanating from the poles of the apparent horizon (white) and those in the disk (green) for all
  cases in Table~\ref{table:Iparamenters}. The arrows indicate the fluid velocities while the BH apparent horizons
  are displayed as black objects. The insets highlight the direction  (not the magnitude)  of the BH spin and
  the field lines once an outflow has been launched.
\label{fig:outcome}}
\end{center}
\end{figure*}

To probe periodicity features in the accretion rate that may give rise to periodic EM signatures,  we perform a
Fourier analysis as in~\cite{Gold:2013zma,Khan:2018}. We find that, regardless the mass ratio, the total
accretion rate $\dot{M}_{\rm BHBH}$ (i.e. the accretion mass onto both BHs) exhibits a dominant frequency
$2\,\pi\,f_{\rm c}\simeq (3/4) \Omega_{\rm BHBH}$. Here $\Omega_{\rm BHBH}$ is the average orbital frequency
determined during the predecoupling phase from the GWs. This characteristic frequency is roughly the same as
that found in our GRMHD studies of circumbinary disks around nonspinning BHBHs with mass ratio $q= 36/29$
in~\cite{Khan:2018}, though slighly smaller than~$2\,\pi\,f_{\rm c}\simeq (4/3)\Omega_{\rm
  BHBH}$ found in our GRMHD studies of equal-mass, nonspinning BHBHs in~\cite{Farris:2012,Gold:2013zma},
where it has been attributed to the Lindblad resonance~\cite{Macfadyen:2006jx}. In addition, the GRMHD studies
in~\cite{Gold:2013zma}, where an effective radiative cooling process was also considered, found a single peak frequency
at  $\Omega_{\rm BHBH}$ in $q=4$. Our results indicate that periodicity signatures in the accretion rate during
the predecoupling phase are not sensitive to the binary mass ratio, to the BH spin, or to the geometric thickness
of the disk. The latter is suggested in~\cite{Khan:2018}.  These results may be used potentially to infer the binary
frequency from EM observations when the GWs are too weak, though realistic cooling processes are required to reliably determine how  the
variability in the accretion rate maps onto EM signatures. The Newtonian simulations in~\cite{2013MNRAS.436.2997D,
  Farris:2014} of a merging BHBH in a geometrically-thin disk with mass ratio~$q\gtrsim 10$ and an $\alpha$-viscosity
prescription found that the characteristic frequency of the accretion rate strongly depends on the mass ratio.
The binary induces eccentricity in the inner region of the disk, driving the formation of overdense lumps which 
enhance periodicities patterns in the accretion rate. This effect may provide a method for observationally inferring
mass ratios from luminosity measurements that give the inferred accretion rate.  However, GRMHD simulations of BHBH
in geometrically thick disks found a rather complex structure in the Fourier spectrum of $\dot{M}$~\cite{Gold:2013zma}.
These differences have been attributed mainly to the different turbulent viscosity prescriptions.

Once the system reaches the  postdecoupling phase (roughly after~$11$~quasi-circular orbits; see bottom panel in
Fig.~\ref{fig:Mdot_GW}),  the accretion rate gradually decreases as the BHBH inspiral speeds up until merger.
Following merger, low-density material refills the partial hollow left by the binary causing the accretion to ramp
up once again. As the spinning BH remnant reaches a quasi-steady state, the accretion for $q=1$ settles to $\sim 40\%$
of the maximum value of the rate reached during the binary inspiral, while it settles at $\sim 43\%$ and $\sim 48\%$
of the maximum value of the accretion rate for $q=2$ and $q=4$, respectively.  The trend is that
the smaller the spin of the BH remnant (the larger the ISCO; see Table~\ref{table:results}), or equivalently,
the larger the binary mass ratio of the progenitors, the larger the relative accretion rate.
%
\begin{figure*}
\begin{center}
\includegraphics[width=2.2\columnwidth]{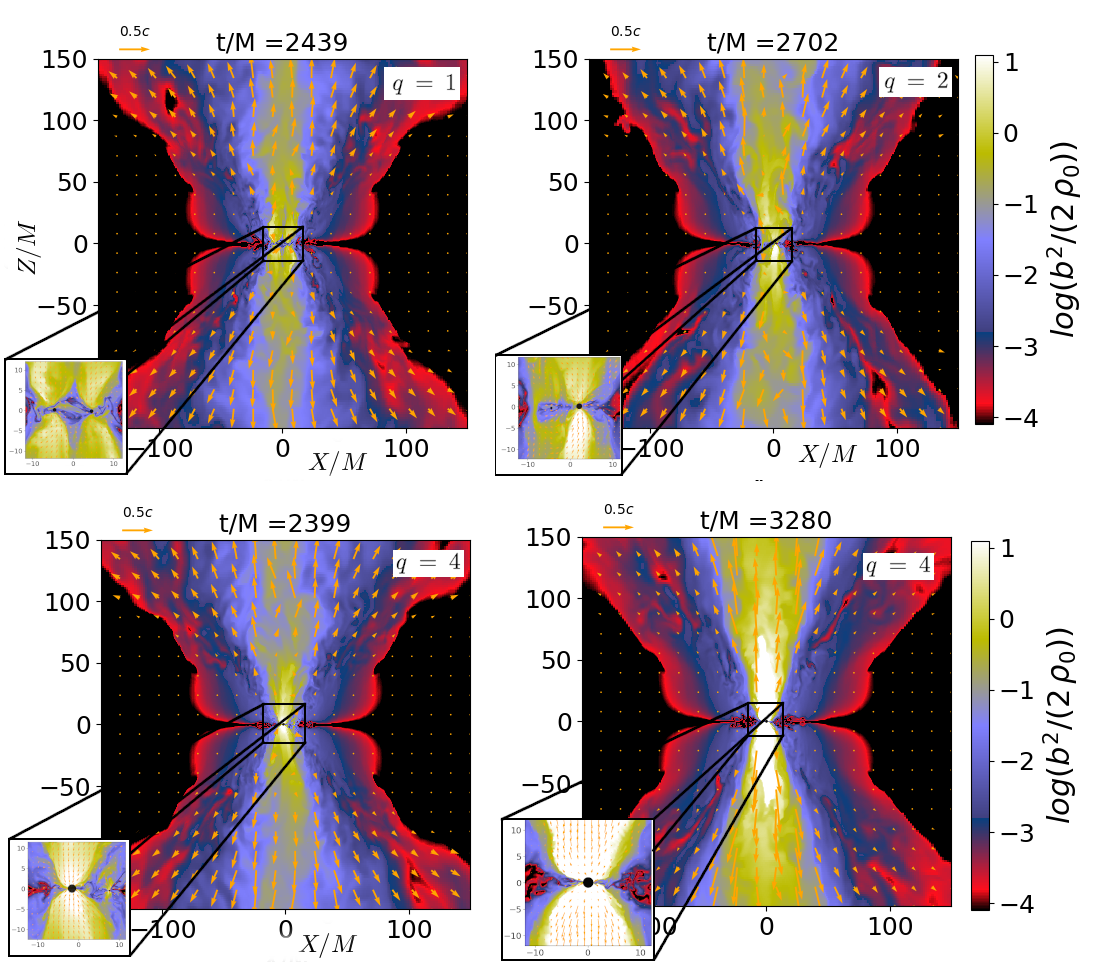}
\caption{Force-free parameter $b^2/(2\rho_0)$ (log scale) on a meridional plane during the predecoupling
  phase for all cases in Table~\ref{table:Iparamenters} (top and bottom-left panels) and during the
  postmeger rebrightening/afterglow phase for $q=4$ case (bottom-right panel). Similar results are found for
  the other cases considered here. The arrows indicate the fluid velocities while the BH apparent horizons
  are displayed as black objects. The inset highlights the force-free parameter in the vicinity of the BH
  apparent horizons.
}
\label{fig:outflow}
\end{center}
\end{figure*}

Anisotropic emission of GWs and  momentum conservation produces a recoil or ``kick'' of the BH remnant, which
can have a significant impact on the accretion flow. However, in our $q=2$ and $q=4$ cases the kick velocity
is $\lesssim 230\,\rm{km/s}$ (see Table~\ref{table:results}), and hence it is small compared to other
characteristic velocities in the system~\cite{Bogdanovic:2007hp}. It is expected that these kicks do not
have any significant impact on the accretion.
%
%
\begin{center}
  \begin{table}[th]
    \caption{Summary of main results. Columns display case name, merger time defined as the time of GWs peak
      amplitude. Followed by the mass, the spin and the kick velocity of the BH remnant in $\rm km\,s^{-1}$. Next, the time-averaged
      accretion rate onto the BH remnant
      over the last $\Delta t\sim 500M$ before the termination of our simulations,
      the Poynting radiative efficiency $\epsilon_{\rm EM}=L_{\rm EM}/\dot{M}$ and the Lorentz factor in
      the funnel after the outflow has reached quasi-equilibrium following merger.
      \label{table:results}}
    \begin{tabular}{cccccccc} 
      \hline
      \hline
      case & $t_{\rm mer}$ &$M_{\rm BH}$  &$S_z/M_{\rm irr}^2$ & $v_{\rm kick}$  & $\dot{M}/\dot{M}_{\rm eq}^\dag$ & $\epsilon_{\rm EM}$ & W\\
      \hline
      q1   & 3425 & $0.94M$      &            0.78    &  3       & $1.23$ & 0.030 & 2.6  \\
      q2   & 2962 & $0.96M$      &            0.74    &  230     & $1.30$ & 0.026 & 2.1  \\
      q4   & 3148 & $0.98M$      &            0.60    &  170     & $1.38$ & 0.024 & 1.6  \\
      \hline
    \end{tabular}
    \begin{flushleft}
      $^{(\dag)}$ $\dot{M}_{\rm eq}\sim 10^{-2}
      (\frac{\rho_0}{10^{-11}\rm gm/cm^{3}})\,(\frac{M_{\rm BH}}{10^6M_\odot})^2\,M_\odot\,\rm yr^{-1}$.\\
    \end{flushleft}
  \end{table}
\end{center}

\section{Outflows and jets}
Fig.~\ref{fig:outcome} displays our 3D visualizations of the rest-mass density along with magnetic field
lines during the late predecoupling (left column) and rebrightening/afterglow (right column) phases.
During the former (i.e.~once the accretion reaches a quasi-stationary state), the magnetic pressure above
the poles of each individual BH increases as net poloidal magnetic fluxes are accreted onto them. This in
turn leads to the formation of a helical funnel bounded by the magnetic field lines emanating  from the BHs.
These funnels lie along the direction of the BHs' spin and form as long as the force-free parameter is $b^2/(2\rho_0)
\gtrsim 1$ (i.e. along magnetically-dominated regions). However, at larger distances from BH poles,
where the magnetic-to-gas-pressure ratio decays to values $\lesssim 1$, the funnels are reoriented 
in the direction of the total angular momentum of the system. In particular, the primary BH's
funnel in $q=4$ extends a distance of~$r \sim 10~r_{\rm BH}$ along the direction of its spin and then
is reoriented along the  total angular momentum of the system. See insets on
the left column in Fig.~\ref{fig:outcome} for other cases. This effect leads to a dual jet structure in the
vicinity of the BHs that resembles the one reported in~\cite{Gold:2013zma}. Here $r_{\rm BH}$ is the radius
of the apparent horizon.

Fig.~\ref{fig:outflow} displays the force-free parameter on a merional slice during the predecoupling epoch
(see top and bottom-left panels). We observe that in the vicinity of the primary BHs' poles this parameter
reaches values $\sim 100$ for cases with mass ratio $q> 1$. However, for the secondary BHs and those with
$q=1$ its values reach at most $b^2/(2\rho_0)\sim 10$ (see insets in~Fig.~\ref{fig:outflow}). This behaviour
is anticipated because
the pile up of the matter inside  the Hill sphere of the latter BH maintains a higher density and lower force-free
parameter (see Sec.~\ref{sec:accretion}). The larger values of $b^2/(2\rho_0)$  help
to tighten the magnetic field in the funnel, leading to a more collimated helical structures. This explains
why the primary BH for cases with $q>1$ has a narrower funnel than that above the poles of the secondary BH
(see bottom inset in Fig.~\ref{fig:outcome}). As the magnetic-to-gas-pressure increases, persistent outflows
confined within the BH funnels are launched and merge into a single one at larger distance $r\gg 10\,
r_{\rm BH}$ 
(see Fig.~\ref{fig:outflow}).
As shown on the right column in~Fig.~\ref{fig:outcome}, incipient jets persist during the whole evolution.
Similar results have been observed in~\cite{Cattorini:2022tvx} once the BHBH-cloud of matter reaches a 
quasi-stationary state. As pointed out in~\cite{Gold:2013zma} these results suggest that it is unlikely that
EM signatures from the individual jets can be detectable, as suggested in previous force-free
simulations~\cite{Palenzuela:2010nf}. As we described previously, during the predecoupling phase the binary
carves out a partial hollow where  accretion streams constantly feed the BHs. Therefore, a force-free
prescription everywhere as considered in~\cite{Palenzuela:2010nf} may not be suitable to model this
system.

During merger, a significant fraction of the orbital angular momentum of the binary is transferred to that
of the remnant BH causing an appreciable increase in the magnitude and a sudden reorientation of its spin with
respect to those spins of the progenitors (see~Fig.~\ref{fig:SpinBHs}). Simultaneously, the two funnels merge into
a single tightly wound, collimated, helical magnetic funnel pointing along the direction of the BH spin remnant.
As the latter spin increases by a factor $\gtrsim 2$  with respect to those of the progenitor
(see~Table~\ref{table:results}), and the inner disk's edge flows inward, the outflow collimates even further.
During this period we observe that magnetically-dominated regions with values~$b^2/(2\rho_0)\gtrsim 100$ rapidly
expand to heights larger than $r~\sim 100\,r_{\rm BH}$
(see bottom-right panel in Fig.~\ref{fig:outflow}). This is a typical effect of the BH spin on the magnetic
field lines that has been observed in differerent astrophysical scenarios involving compact binary mergers
(see~e.g.~\cite{Gold:2014b,Cattorini:2022tvx,Kelly:2017,Palenzuela:2010nf,Baiotti:2016qnr,Ruiz:2021gsv}).

After merger, the sudden reorientation of the BH spin, and hence of the magnetic field lines, induces the formation
of a kink which propagates along the helical magnetic funnel. We track its propagation using our 3D
visualizations of the magnetic field lines every $\Delta t\simeq 0.5M$ and find that, regardless the
mass ratio, the kink is damped in $\Delta t \lesssim 15M$ following its formation, leaving no memory
of the reorientation of the BH spins. 
Similar behavior can be inferred from other misaligned studies (see e.g. bottom-right
panel in Fig.~2. in~\cite{Cattorini:2022tvx}). These results suggest that X-shaped radio galaxies cannot
be formed  through galaxy mergers  containing misaligned, low-spin ($\chi \lesssim 0.6$) SMBHs in low-mass disks
and hence rule out the spin-shifters as an unique mechanism to explain X-shaped radio galaxies if the accretion
onto the BH is stochastic as previously claimed~(see e.g.~\cite{King:2008au,2012ApJ...753...15N}). Other events
may therefore be necessary to induce helical distortions leading to an X-shaped jet.
For example, the interactions between the jet
and the shells of stars and debris that form and rotate around the merged galaxy may cause temporary deflections
of the jets, creating the observe of structure~\cite{Gopal-Krishna:2010qxl}. Alternatively, precession of a
highly spinning BH in self-gravitating, massive disk may provide another explanation~\cite{Tsokaros:2022hjk}.

We track the Lorenz factor $W=\alpha_c\,u^0$ of the flow inside regions where~$b^2/(2\rho_0)\geq 10^{-2}$,
our definition of the funnel's boundary. Here $\alpha_c$ is the lapse function. During the predecoupling
phase, and once the accretion rate has reached a quasi-steady state, we observe that the maximum value of the
Lorentz factor depends only slightly on the mass ratio. At a modest distance from the BHs ($r\gtrsim 30 r_{\rm BH}$)
$W \sim 1.1$ for $q=1$ while $W\sim 1.2$ and  $1.3$ for $q=2$ and $q=4$, respectively, and hence the  larger the
mass ratio the larger the Lorentz factor of the mildly relativistic outflow. However, this behavior is reversed
after merger, where is found that $W \sim 2.6$ for $q=1$, $W \sim 2.1$ for $q=2$ and $W \sim 1.6$ for $q=4$.
This relative change is likely  due to the magnitude of the spin of the BH remnant~(see Table~\ref{table:results}).
As mentioned before, the lower the binary mass ratio, the larger the spin of the BH remnant, which
drives a more confined fluid beam within a more collimated, tightly wound, helical magnetic field.
Note that steady-state and axisymmetric jet models show that the maximum attainable value of $W$ at large distance
is approximately equal to $b^2/(2\rho_0)$~\cite{Vlahakis:2003si}, which reaches values $\gtrsim 100$ within
the funnel. Therefore, material in the above funnels may be accelerated to $W\sim 100$, as requiered
to explain $\gamma$-ray burst phenomenology.

\begin{figure}
  \includegraphics[scale=0.32]{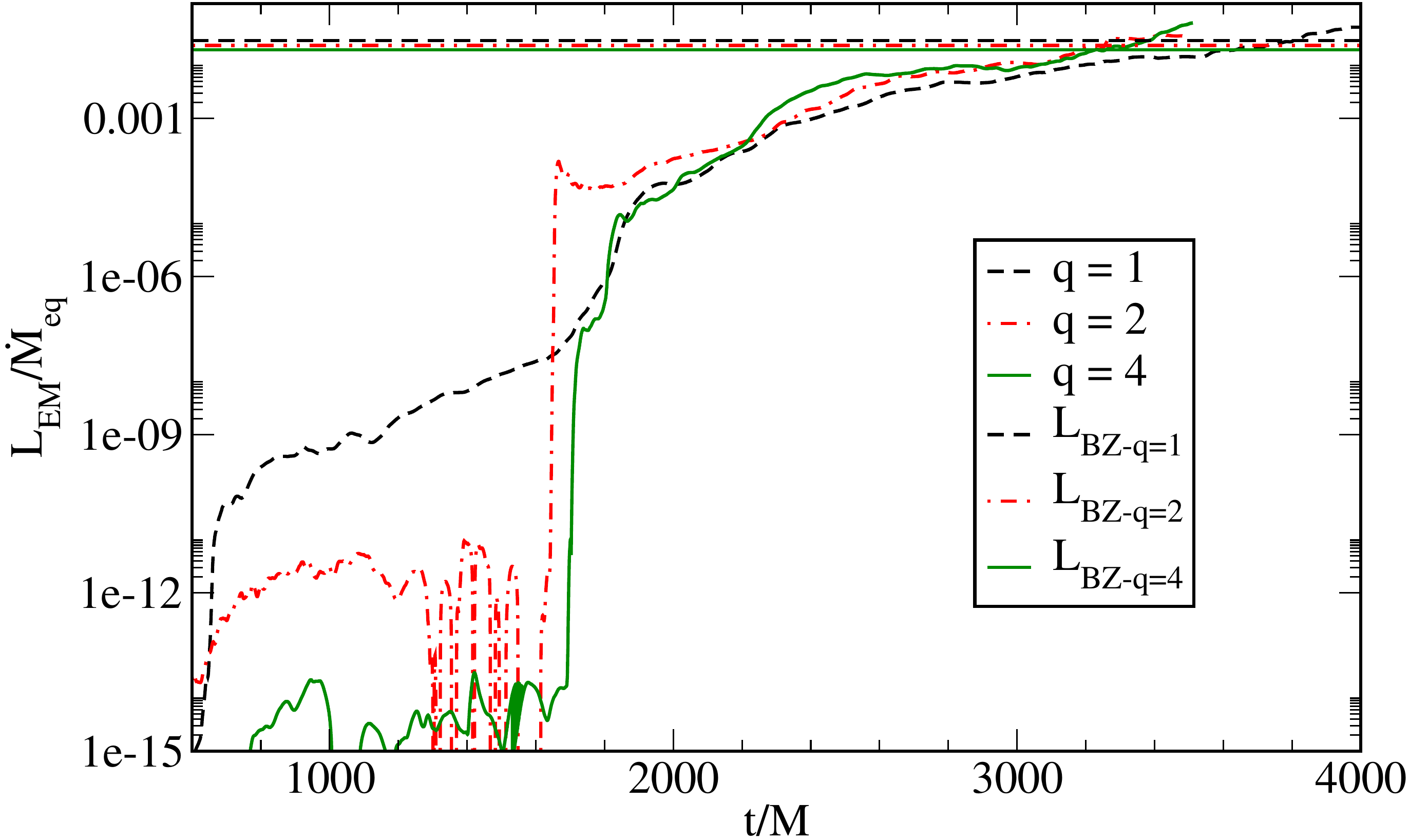}
\caption{
  Poynting luminosity $L_{\rm EM}$ for cases in Table~\ref{table:Iparamenters} normalized
  to the time-averaged accretion rate onto the BH remnant over the last $\Delta t\sim 500M$ before
  the termination of our simulations (i.e. the radiative efficiency).
  The horizontal lines indicate the expected Blandford-Znajek luminosity $L_{\rm BZ}$ from a single BH with the
  same mass, spin and quasi-stationary polar magnetic field strength as those found for the BH-disk remnant
  in each simulated case.
\label{fig:PoyntingLEM}}
\end{figure}

We compute the EM Poynting flux $L_{\rm EM}=-\int T^{r\,\rm (EM)}_t\,\sqrt{-g}\,d\mathcal{S}$
across a spherical surface $\mathcal{S}$ with a coordinate radii ranging between $r_{\rm ext}=120M$
and $r_{\rm ext}=300M$. Here $g$ is the full spacetime metric determinant and $T^{\mu\,\rm (EM)}_\nu$ is 
the EM stress-energy tensor. Fig.~\ref{fig:PoyntingLEM} displays $L_{\rm EM}$ as a function
of time for all cases in Table~\ref{table:Iparamenters}. During the predecoupling phase, we observe that
the EM luminosity in the $q=1$ case increases smoothly during the first $\sim 6$ quasi-circular orbits
($t\sim 1600M$; see bottom panel in Fig.~\ref{fig:Mdot_GW}) followed by a sudden boost of roughly three
orders of magnitude in only $\sim 1.6$ orbits
[$\Delta t\sim 230M\simeq 0.36 (M/10^6M_\odot)(1+z)\,\rm h $ or
  $\Delta t\sim 230M\simeq 14 (M/10M_\odot)(1+z)\,\rm ms$]. After this, $L_{\rm EM}$ smoothly
ramps up reaching a nearly constant value of $L_{\rm EM}\sim \,0.015 \dot{M}_{\rm eq}$ as the accretion
rate reaches a quasi-equilibrium state (see Fig.~\ref{fig:Mdot_GW}). Here $\dot{M}_{\rm eq}$ is the
time-averaged accretion rate onto the BH remnant over the last $\Delta t\sim 500M$ before the termination of our
simulation. Presumably the pre-equilibrium rise is due to the system adjusting to our adopted initial conditions,
where a vacuum exists around the binary.
Similar behavior is observed in $q=4$ though the luminosity increases by a factor of
$\sim 10^2$ in $\sim 1$ orbit [$\Delta t\sim 100M \simeq  0.14 (M/10^6M_\odot)(1+z)\,\rm h $ or
$\Delta t\sim 100M \simeq  5 (M/10M_\odot)(1+z)\,\rm ms$]. After that, it
smoothly increases until it reaches the nearly constant value~$\sim 0.010\dot{M}_{\rm eq}$.
This behavior is not evident in $q=2$ where the luminosity jumps discontinuously from almost
null values to a nearly constant value of~$\sim 0.012 \dot{M}_{\rm eq}$. Once the BH remnant settles, we compute
the efficiency defined as $\epsilon_{\rm EM}\equiv L_{\rm EM}/\dot{M}_{\rm eq}$ and find that
the smaller the mass ratio the larger the $\epsilon_{\rm EM}$
(see Table~\ref{table:results}). This is most likely due to the fact that the
poloidal magnetic flux, which has been found to be one of the major determining factors for the efficiency of the
outgoing EM (Poynting) luminosity in MHD BH-accretion disk systems (see~e.g.\cite{Beckwith:2007sr,
  McKinney:2012vh}), is larger when the binary mass ratio is smaller.

Following  merger, the magnitude of the Poynting luminosity in all cases gets a new, but rather moderate, boost
and then settles to nearly constant value until the termination of our simulations. A more substantial rise likely
occurs in the radiation from the disk, including a thermal component \cite{Shapiro:2013qsa}.
In contrast to the non-spinning GRMHD
BHBH-disk simulations we reported in~\cite{Gold:2014b}, where it was found that during the postmerger phase the boost
in the outgoing Poynting luminosity decreases as the mass ratio increases, we observe that $L_{\rm EM}$ is boosted by
roughly a factor of $\sim 4$ regardless the mass ratio.  These results may indicate that the effects of the BH
spin on this luminosity have a stronger impact during the predecoupling phase. As in~\cite{Gold:2014b}, we also
observe a time delay between the binary merger and the boost which depends on the mass ratio. In particular, we
find that in $q=1$ the increase in the luminosity takes place at $\Delta t \sim 425M\simeq  0.6(M/10^6M_\odot)
(1+z)\,\rm h$ [or $\Delta t \sim 425M\simeq  2.1(M/10M_\odot)(1+z)\,\rm ms$]
after merger, while in $q=4$ it occurs at $\Delta t\sim 160M\simeq 0.2 (M/10^6M_\odot)(1+z)\,
\rm h$ [or $\simeq  0.8(M/10M_\odot)(1+z)\,\rm ms$], and hence the larger the binary mass ratio
the shorter the time delay. A similar effect has been found
by fixing the mass ratio but changing the disk thickness in~\cite{Khan:2018}.

As the BHs are spinning  with nearly force-free fields inside their jet funnels, one may attribute 
the Pointyng luminosity to the Blandford-Znajek mechanism~\cite{Blandford:1977}, which is typically invoked to
describe the emission from the force-free magnetospheres above the poles of 
single, spinning BHs~(see e.g.~\cite{McKinney:2004ka}). Following merger, we compute the expected
Blandford-Znajek EM luminosity $L_{\rm BZ}$ (see~Eq.~4.50 in~\cite{Thorne86}) from a single BH with the same mass
and spin parameter $\chi$ as the remnants in the postmerger phase of our simulations~(see~Table~\ref{table:results})
and the quasi-stationary polar magnetic field strength time-averaged over the last $\Delta t\sim 500M$ before the
termination of our simulations: $L_{\rm BZ}\sim 10^{51}\,\chi^2\,(M_{\rm BH}/10^6M_\odot)^2\,(B/10^5G)^2{\rm erg\,
  s^{-1}}$ [or  $L_{\rm BZ}\sim 10^{51}\,\chi^2\,(M_{\rm BH}/10M_\odot)^2\,(B/10^{16}G)^2{\rm erg\,
  s^{-1}}$].
As displayed in Fig.~\ref{fig:PoyntingLEM}, $L_{\rm BZ}$ is roughly consistent with the $L_{\rm EM}$ extracted
from our simulations and hence we can attribute this luminosity and accompanying magnetically-power
jet from the spinning BH-disk remnant system to the Blandford-Znajek mechanism. This value is also consistent with the
narrow range of values predicted  for the jet luminosity for {\it all} such as systems, independent of mass scale \cite{Shapiro:2017cny}.
%
%
%
%
\section{Discussion}
\label{sec:discussion}
The advance of ``multimessenger astronomy'' promises to resolve a number of long-standing astrophysical puzzles,
such as the origin of $\gamma$-ray bursts, the nature of matter at supranuclear densities, the formation channels
of supermassive black holes, etc. However, to understand these observations and, specifically
to understand the interplay between general relativity, GW and EM signals, and the underlying
microphysics, it is crucial to compare observations to predictions from theoretical modeling.
In particular, multimessenger observations of accreting black holes, single and binaries,
call for a detailed understanding of the environment surrounding them and the knowledge of some
characteristic GW and EM signatures that can be used to distinguish these systems from other compact
objects and determining their parameters.

As a step forward in this goal we extended our previous GRMHD simulations of
binary black holes in magnetized, circumbinary disks~\cite{Farris:2012,Gold:2013zma,Gold:2014b,Khan:2018}
by considering misaligned, spinning binary black holes undergoing merger with mass ratios $q=M_{\rm 1,irr}/
M_{\rm 2,irr}=1$, $2$ and $4$. The magnitude of the individual, dimensionless, black hole spin is $\chi\equiv
S/M^2_{\rm i,irr} = 0.26$, consistent with a stochastic accretion in AGN~(see e.g.~\cite{King:2008au,2012ApJ...753...15N}),
lying either along the initial orbital plane or $45^\circ$ above it.

We evolved these configurations starting from the end of the predecoupling phase all the way to the merger and
rebrightening/afterglow phase to  identify their Poynting EM emission features and to analyze their dependence on the black hole
spin and the binary mass ratio. We also discussed whether the spin-shift of the black hole remnant with respect
to those of the progenitors leaves an observable imprint in the outgoing Poynting luminosity, surrounding medium profile
or in the magnetically driven-jet that can be used to characterize the spin properties of merging black hole binaries
or give new insight into the formation channels of X-shaped radio galaxies~\cite{1984MNRAS.210..929L,Gopal-Krishna:2003bjt}
and searches for such systems (see e.g.~\cite{2015ApJ...810L...6R}).

During the predecoupling phase, we observed that a magnetically-driven (dual) outflow emerges from the poles of
the two black holes, pointing along the direction of their spin as long as the force-free parameter within
the funnel satisfies $b^2/(2\rho_0)=B^2/ (8\,\pi\rho_0)\gtrsim 1$. At large distance, where the values of this
parameter drops to $\lesssim 1$, the twin jets merge, generating a single outflow. This effect
may prevent the EM detection of the individual jets, as has been suggested previously~\cite{Palenzuela:2010xn}.
During this phase, we also found that  the accretion rate exhibits a quasi-periodic behavior with a dominant
frequency $2\,\pi\,f_{\rm c}\simeq (3/4)\Omega_{\rm BHBH}$ that is mass ratio independent. Here
$\Omega_{\rm BHBH}$ is the average orbital frequency of the binary. Simultaneously, we observed that, depending
on the mass ratio, the outgoing Poynting luminosity is significantly boosted during $\sim 0.14 -0.36 (M/10^6
M_\odot)(1+z)\,\rm h$ [or $\sim 5 -14 (M/10M_\odot)(1+z)\,\rm ms$] after the start of the simulations, though
this may be an artifact of our initial
conditions, which are characterized, by a vacuum close to the binary. Consistent with~\cite{Gold:2014b}, and after
quasi-equilibrium accretion is achieved, we observed a new, but moderate, boost in the luminosity during
the re-brightening/afterglow phase following meger that is nearly mass ratio independent. These sudden changes 
can alter the Poynting emission across the jet and may be used to
distinguish black hole binaries in AGNs from single accreting black holes based only on jet morphology in
NGC~5128-like sources~\cite{Gold:2014b}.

During and after merger, we tracked the field lines emanating from the black hole apparent horizon and observed a
kink perturbation propagating along the funnel. The kink can be attributed to the spin reorientation of the black hole
remnant with respect to the spins of the progenitors. Using our 3D visualizations every $\Delta\sim 0.5 M$, we
followed this perturbation and found that it is quickly damped in a  few $M$, leaving no memory of the spin
reorientation. Similar results were inferred from the  misaligned spinning black hole binary immersed in a
magnetized cloud of matter~\cite{Cattorini:2022tvx}. These results disfavor the spin reorientation mechanism
following binary black hole mergers as a plausible scenario to explain X-shaped radio galaxies, at least in low-mass
disks and low-spin black
holes. Note that if stochastic accretion onto supermassive black holes  is taking place, which naturally induces
low spinning BHs, then additional mechanisms, such as jet interaction with the
environment~\cite{Gopal-Krishna:2010qxl}, remnant black hole precession, due to tidal torques and/or gravitomagnetic
precession in a more massive ambient disks, may be required to explain this
scenario.

\begin{acknowledgments}
  We thank members of our Illinois Relativity Undergraduate Research Team (M. Kotak, J. Huang, E. Yu, and J. Zhou)
  for assistance with some of the 3D visualizations.  This work was supported in part by National Science Foundation
  (NSF) Grant PHY-2006066 and the National Aeronautics and Space Administration (NASA) Grant 80NSSC17K0070 to the
  University of Illinois at Urbana-Champaign. M.R. acknowledges support by the Generalitat Valenciana Grant
  CIDEGENT/2021/046 and by the Spanish Agencia Estatal de Investigaci\'on (Grant PID2021-125485NB-C21). 
  A.T. acknowledges support from the National Center for Supercomputing Applications (NCSA) at the University of
  Illinois at Urbana-Champaign through the NCSA Fellows program. This work made use of the Extreme Science and
  Engineering Discovery Environment (XSEDE), which is supported by NSF Grant TG-MCA99S008. This research is also
  part of the Frontera computing project at the Texas Advanced Computing Center.
  Frontera is made possible by NSF award OAC-1818253.
   Resources supporting this work were  also provided by the NASA High-End
   Computing Program through the NASA Advanced Supercomputing Division at Ames
   Research Center. The authors thankfully acknowledge the computer resources at
   MareNostrum and the technical support provided by the Barcelona Supercomputing Center
   (AECT-2023-1-0006).
\end{acknowledgments}

\bibliographystyle{apsrev4-1}
\bibliography{ref}

\end{document}